\documentclass[12pt,preprint]{aastex}

\newcommand{\feh}{\mbox{${\rm[Fe/H]}$}}

\newcommand\teff{{T_{\rm eff}}}

\newcommand\lta{\mathrel{\hbox{\raise 0.6 ex \hbox{$<$}\kern
                   -1.8 ex\lower .5 ex\hbox{$\sim$}}}}
\newcommand\gta{\mathrel{\hbox{\raise 0.6 ex \hbox{$>$}\kern
                   -1.7 ex\lower .5 ex\hbox{$\sim$}}}}
 
\shortauthors{VandenBerg et al.}
\shorttitle{Stellar Models for Different Metal Mixes}
 
\begin{document}
 
\title{Models for Metal-Poor Stars with Enhanced Abundances of C, N, O, Ne, Na,
Mg, Si, S, Ca, and Ti, in Turn, at Constant Helium and Iron Abundances}

\author{Don A.~VandenBerg}
\affil{Department of Physics \& Astronomy, University of Victoria,
       P.O.~Box 3055, Victoria, B.C., V8W~3P6, Canada}
\email{vandenbe@uvic.ca}

\author{Peter A.~Bergbusch}
\affil{Department of Physics, University of Regina, Regina, Saskatchewan,
       S4S 0A2, Canada}
\email{pbergbusch@accesscomm.ca}

\author{Aaron Dotter\altaffilmark{1}}
\affil{Department of Physics \& Astronomy, University of Victoria,
       P.O.~Box 3055, Victoria, B.C., V8W~3P6, Canada}
\email{Aaron.Dotter@gmail.com}

\author{Jason W.~Ferguson}
\affil{Department of Physics, Wichita State University, Wichita KS 67260-0032,
       U.S.A.}
\email{Jason.Ferguson@wichita.edu}

\author{Georges Michaud}
\affil{D\'epartement de Physique, Universit\'e de Montr\'eal, Montr\'eal,
       Quebec, H3C 3J7, Canada}
\email{michaudg@astro.umontreal.ca}

\author{Jacques Richer}
\affil{D\'epartement de Physique, Universit\'e de Montr\'eal, Montr\'eal,
       Quebec, H3C 3J7, Canada}
\email{jacques.richer@umontreal.ca}

\author{Charles R.~Proffitt}
\affil{Space Telescope Science Institute, 3700 San Martin Drive, Baltimore,
       MD 21218, U.S.A.}
\email{proffitt@stsci.edu}

\altaffiltext{1}{Current Address: Space Telescope Science Institute,
      3700 San Martin Drive, Baltimore, MD 21218, U.S.A.}

\begin{abstract}
Recent work has shown that most globular clusters have at least two
chemically distinct components, as well as cluster-to-cluster differences
in the mean [O/Fe], [Mg/Fe], and [Si/Fe] ratios at similar [Fe/H] values.
In order to investigate the implications of variations in the
abundances of these and other metals for H-R diagrams and predicted ages,
grids of evolutionary sequences have been computed for scaled solar and
enhanced $\alpha$-element metal abundances, and for mixtures in which the
assumed [$m$/Fe] value for each of the metals C, N, O, Ne, Na, Mg, Si, S,
Ca, and Ti has been increased, in turn, by 0.4 dex {\it at constant} [Fe/H].
These tracks, together with isochrones for ages from $\approx 6$ to 14 Gyr, 
have been computed for $-3.0 \le$ [Fe/H] $\le -0.6$, with helium abundances
$Y = 0.25$, 0.29, and 0.33 at each [Fe/H] value, using upgraded versions of
the Victoria stellar structure program and the Regina interpolation code,
respectively.  Turnoff luminosity versus age relations from isochrones
are found to depend almost entirely on the importance of the CNO-cycle,
and thereby mainly on the abundance of oxygen.  Since C, N, and O, as well
as Ne and S, do not contribute significantly to the opacities at low
temperatures and densities, variations in their abundances do not impact
the predicted $\teff$ scale of red giants.  The latter is a strong function
of the abundances of only Mg and Si (and Fe, possibly to a lesser extent),
because they are so abundant and because they are strong sources of opacity
at low temperatures.  For these reasons, Mg and Si also have important
effects on the temperatures of MS stars.  Due to their low abundances, Na,
Ca, are Ti are of little consequence for stellar models.  The effects of
varying the adopted solar metals mix and the helium abundance at a fixed
[Fe/H] are also briefly discussed.
\end{abstract}

 
\section{Introduction}
\label{sec:intro}
 
It is now a well-established result that the observed chemical abundance
variations in globular cluster (GC) stars are due to both evolutionary processes
within them and variations in the chemical makeup of the gas out of which they
formed.  There is no doubt that extra nonconvective mixing (not yet fully
understood, but see \citealt{den12} for recent advances in our understanding)
operates in cluster giants brighter than the red-giant-branch (RGB)
bump (\citealt{dv03}, and references therein), which causes the surface
abundances of C, N, and (sometimes) O to vary with luminosity.  However,
CN variations and the O--Na anticorrelation that appears to be characteristic
of nearly all GCs (\citealt{cbg09}) have also been found in dwarf, turnoff,
and subgiant stars (e.g., \citealt{ccb98}; \citealt{gbb01}; \citealt{rc02};
\citealt{cm05}; \citealt{dlg10}), where extra mixing is not a viable
explanation.  (NGC$\,$5466 appears to be a notable exception to this insofar
as little or no evidence has been found in this system for primordial
star-to-star variations in the abundances of the light elements; see
\citealt{smw10}.)  Indeed, below the RGB bump, the chemical composition
anomalies do not appear to vary with luminosity (also see \citealt{gbn02}).

Further evidence for primordial abundance variations in GCs is provided by the
detection of Mg--Al anticorrelations (see \citealt{ygn05}).  The stars currently
on the main sequence (MS) and the RGB of these systems can hardly be responsible
for the Al-rich, Mg-depleted (and, in a subset of the latter, Si-enhanced) stars
that have been found in the most massive and/or the most metal-deficient
clusters (like NGC$\,$2808, NGC$\,$6388, and M$\,$15 --- see \citealt{cbg09})
because their interiors are not hot enough for the Mg--Al cycle (and,
especially, any additional processing to Si) to occur.  As discussed by
Carretta et al., the most likely explanation for the origin of the observed Mg,
Al, and Si abundances is hot-bottom burning in intermediate-mass
asymptotic-giant-branch (AGB) stars.  In fact, there are cluster-to-cluster
differences in the mean abundances of these elements as well as star-to-star
differences within a given GC.  For instance, as reported by Carretta et al.,
NGC$\,$2808, M$\,$4, and M$\,$5, which have similar [Fe/H] values to within
$\approx 0.3$ dex, have mean [Mg/Fe] ratios of 0.20, 0.41, and 0.55, with
{\it rms} variations of 0.25, 0.07, and 0.03 dex, respectively, while their
[Si/Fe] values are, in turn, 0.28, 0.30, and 0.52, with {\it rms} variations
amounting to $\approx 0.05$ dex in each cluster.  Interestingly, the average
[O/Fe] value in the Carretta et al.~sample of 17 GCs ranges from $-0.30$ in
NGC$\,$6388 to $+0.46$ in NGC 7099, with {\it rms} variations from $\sim 0.1$
dex in several systems to as high as 0.36 dex (in NGC$\,$2808).

These results raise at least two obvious questions.  Is the net effect of the
scatter in the [$m$/Fe] value for each metal $m$, within a given cluster, large
enough to cause a detectable spread in the effective temperatures, and hence the
colors, of stars along the principal sequences that define its color-magnitude
diagram (CMD)?  Are cluster-to-cluster differences in the abundances of
individual metals big enough to affect the difference in color between the
turnoff and the lower RGB, which is a sensitive measure of the relative ages of
GCs having very similar [Fe/H] values (\citealt{vbs90}) and often used for that
purpose (e.g., \citealt{map09})?  \citet{vs91} showed early on that this
relative age diagnostic is affected by differences in [O/Fe], but it is not yet
clear whether it is also impacted by variations in the abundances of other
$\alpha$-elements, such as Ne, Mg, and Si, which are the most abundant metals
after the CNO group.

Although there have been a few studies of metal-poor stellar models that allow
for C--N--O--Na anticorrelations (notably \citealt{swf06}, \citealt{csp08},
\citealt{pcs09}), with some speculative discussion of the expected effects of
Mg--Al anticorrelations (see the paper by Salaris et al.), the consequences of
varying each of the most abundant metals, in turn, have yet to be adequately
investigated.  To be sure, we already have a very good understanding of the role
played by C, N, and O in the evolution of lower mass stars and on computed
isochrones (e.g., \citealt{rc85}, \citealt{van85},
\citealt{van92}\footnote{\citet{dcf07} incorrectly asserted that the models
reported by VandenBerg (1992) ``did not account for enhanced oxygen in the
opacities".  In fact, opacity tables for temperatures $> 1$ eV ($\approx
12,000$ K) were derived for the assumed heavy-element abundances from the Los
Alamos Astrophysical Opacity Library (\citealt{hmm77}), from which Rosseland
mean opacities could be calculated for any mixture of the 20 most abundant
elements.  It was only the opacities for lower temperatures that were not
obtainable at that time, but as mentioned therein and confirmed in the present
investigation, oxygen does not contribute significantly to the low-$T$ opacities
because it has a high ionization potential and thus is
a poor electron donor at the temperatures and densities
characteristic of the outer layers of stars (also see \citealt{bo86},
\citealt{vb01}).}).  The same can be said for the case when the abundances of
all of the $\alpha$-elements vary together (e.g., \citealt{scs93},
\citealt{vsr00}, \citealt{kdy02}, \citealt{pcs06}).  However, relatively little
has been done to date to examine the dependence of model $\teff$s on the
abundances of individual $\alpha$-elements heavier than oxygen.

The most noteworthy of the few available studies that have carried out such
work is that by \citet[also see \citealt{lwd09}, who discuss some of the 
implications of the Dotter et al.~results for integrated colors, Lick indices,
and synthetic specta]{dcf07}, but their analysis is
complicated by the fact that their computations were carried out at a constant
value of $Z$ (the mass-fraction abundance of all elements heavier than helium).
This has the consequence that, when the abundance of one metal is increased,
the abundances of all of the others are reduced.  The net effect on stellar
models therefore depends on how the surface and interior opacities have changed
as a result of varying all of the elemental abundances at the same time, by
different amounts.  Fortunately, the problem is not as intractable as these
few remarks suggest, because the [Fe/H] values that are obtained for the cases
when the abundances of most of the metals (except oxygen) are enhanced,
in turn, differ by $\lta 0.05$ dex from that of the base mixture.  Hence, the
effects of varying the chemical composition at a constant $Z$ is quite a good
approximation to those derived when a single metal is enhanced at a constant
value of [Fe/H].  Accordingly, the authors are able to confidently predict, for
instance, that increased abundances of Mg and Si will result in cooler tracks
and isochrones.

Still, since observers use number abundance ratios to describe the chemical
compositions of stars, stellar models that vary the abundances of a single
element at a constant [Fe/H] value are much more straightforward to interpret.
The latter approach, which is taken in the present study, has the advantage that
the effects of altering the abundances of individual metals with can be
accurately and precisely quantified.  Moreover, whereas \citet{dcf07} considered
only one (high) value of $Z$, we have generated large grids of tracks and
isochrones for [Fe/H] values ranging from $-3.0$ to $-0.6$, in steps of 0.2 dex,
on the assumption of $Y = 0.25$, 0.29, and 0.33 at each adopted iron abundance.
At the lowest metallicities, one can anticipate that variations in the
abundances of elements heavier than oxygen will have little or no impact on
stellar evolutionary computations, as the interior opacity at low $Z$ is due
primarily to bound-free and free-free processes involving H and He.  The CNO
elements, which are also considered here, differ from the heavier metals in
that they affect the nucleosynthesis of hydrogen to helium, and thereby the
structures of stars: an increase in the importance of the CNO-cycle due to
enhanced abundances of these elements will make H-burning more centrally
concentrated because of the high temperature sensitivity of these reactions
compared to the $pp$-chain, and it is well known that the rate at which the
radius of a low-mass star grows with time after leaving the zero-age main
sequence (ZAMS) depends upon the rate at which its central concentration
increases.  These effects can be expected to be important at any metallicity.

\section{The Heavy-Element Mixtures for Which Stellar Models have been Computed}
\label{sec:zmix}

Table~\ref{tab:tab1} lists all of the heavy elements that are considered when
requesting opacities for stellar interior conditions using the Livermore
Laboratory OPAL opacity website\footnote{http://opalopacity.llnl.gov}, along
with three different determinations of their abundances in the Sun.  The most
recent of these is by \citet{ags09}, which updated the findings that were
published by \citet{ags05}.  As shown in Table~\ref{tab:tab1}, the CNO
abundances that were derived in these two studies differ by $\le 0.05$ dex,
while the differences are generally in the range 0.05--0.10 dex for the
heavier metals.  Interestingly, neither of these results has found favor in
the asteroseismology community (see, e.g., \citealt{pd09}, \citealt{sbf09}).
While the inclusion of additional physics not normally considered in Standard
Solar Models (SSMs) may help to reduce the discrepancies between the predicted
and measured oscillation frequencies and improve the fit to the radial variation
of the sound speed, when ``Asplund" solar abundances are assumed in the models
(e.g., \citealt{cmr11}, \citealt{bll11}), SSMs that adopt the solar abundances
given by \citet{gs98} are still the least problematic ones.

In view of this ongoing controversy, we have chosen to generate grids of
evolutionary tracks for the GS98 and the AGS5 cases, since it is of some
interest to examine the impact that the adopted solar mix of metals has on
isochrones for low values of [Fe/H].  As noted above, the differences between
the solar $\log\,N$ values determined by AGS5 and A09 are quite small;
consequently, the main results of this study will be essentially independent
of this choice.  Nevertheless, grids of models for the A09 abundances will be
provided in a forthcoming paper.

Since the focus of this work is on metal-poor stars, an $\alpha$-enhanced
mix of heavy elements should be assumed in the reference models against which
the computations for single-element enhancements are compared.  
Rather than simply adopting a constant 0.3 or 0.4 dex increase in the abundance
of each $\alpha$-element, the mean [$m$/Fe] values that were determined by
\citet{cds04} in their ``First Stars" ESO Large Programme have been adopted:
they are listed in the fifth column of Table~\ref{tab:tab1}.  (Although these
results strictly apply to stars with [Fe/H] values $\lta -2.0$, very similar
results have, for the most part, been found at higher metallicities --- see,
e.g., \citealt{ygn05}.)  Note that the Cayrel et al.~determinations of [$m$/Fe]
for Cr and Mn have also been assumed, though this choice will not affect the
computed stellar models because the abundances of these elements are so low in
the metallicity regime under consideration.  Be that as it may, additional grids
of tracks were computed in which the GS98 and AGS5 $\log N$ abundances were
adjusted by the $\Delta\log N$ values given in the fifth column of
Table~\ref{tab:tab1} and then scaled to the desired [Fe/H] values.  These sets
of models were given the names GSCX and AGSC, respectively, where the ``C" 
indicates ``Cayrel" et al.~[$m$/Fe] abundance ratios.  (Note that the GSCX
models are the same ones that were used in the analyses of $BV(RI)_CJK_S$
photometry of field and GC stars by \citealt{vcs10} and \citealt{bsv10}.)

Finally, additional model grids were computed for the same ranges of [Fe/H] 
and $Y$ noted above in which the abundance of C, N, O, Ne, Na, Mg, Si, S, Ca,
and Ti was increased, in turn, by 0.4 dex over the AGSC abundances.  The names
that were assigned to these computations are listed in the last column of
Table~\ref{tab:tab1}.  To distinguish the grids for different helium and iron
abundances, we have adopted file names of the form ``nnnnyijmkl", where ``nnnn"
is the 4-letter designation for the assumed mixture (e.g., GS98, AGSC, AGxO),
``yij" specifies the adopted helium abundance (e.g., ``y25" represents
$Y=0.25$; the other choices are ``y29" and ``y33"), and ``mkl" indicates the
assumed [Fe/H] value such that, for instance, ``m30" and ``m06" refer to [Fe/H]
$= -3.0$ and $-0.6$, respectively.   Argon is the only $\alpha$-element that
was not subjected to the same analysis as the others: its effects on tracks and
isochrones should be similar to those for neon, but at a much reduced level,
given that it has a lower abundance by $\sim 1.6$ dex.  Aside from C and N, the
only element not a member of the $\alpha$-group for which stellar models were
generated with and without an enhanced abundance is Na.

\section{The Victoria Stellar Evolution Code}
\label{sec:code}

A number of important improvements have been made to the Victoria stellar
structure and evolution code since it was last used for extensive grid
computations (\citealt{vbd06}).  The most significant one is the inclusion of
the diffusion of hydrogen, helium, and the light elements $^6$Li, $^7$Li, and
$^9$Be (but not the heavier metals) using the numerical methods that are
described in considerable detail by \citet[see the Appendices of this
paper]{pm91}, along with the revised treatment of thermal diffusion reported by
\citet{pv91}.  [The LiBeB elements are treated only to constrain the extra
mixing that is needed to limit the efficiency of gravitational settling from
the surface layers of stars --- see below.  Spectroscopic studies have found
either no detectable difference in the surface abundances of Population II
stars between the turnoff and the lower RGB (e.g., \citealt{gbb01},
\citealt{rc02}) or a small variation, amounting to $\lta 0.15$ dex, which can
be made consistent with the predictions of stellar models if some additional
(turbulent) mixing just below surface convection zones is taken into account
(\citealt{kgr06}).  The models computed for the Korn et al.~study also included
radiative accelerations: as emphasized by \citet{rmr02}, ``it is a better
approximation not to let Fe diffuse at all in a $0.8 {{\cal M}_\odot}$ star
than to calculate its gravitational settling without including the effects of
$g_{\rm rad}$(Fe)".  Radiative accelerations have also been found to have
interesting ramifications for the chemical abundance profiles in the vicinity
of the H-burning shell in low-mass red giants (see \citealt{mrr10}).]

Worth mentioning is the fact that the chemical abundance changes due to nuclear
burning, convective mixing, turbulent transport, and ionic diffusion are
considered simultaneously.  Indeed, because of the transport terms, it is
necessary to solve the relevant equations for all of the species in all zones
at the same time, using a layer-by-layer Newton-Raphson scheme similar to that
employed to solve the stellar structure equations (e.g., \citealt{hfg64}).  (Any
surface and central convective regions that exist are treated as if they were
single, uniformly mixed zones, with mass-averaged initial abundances and nuclear
reaction rates.)  Since the solution to the chemical composition equations
affects the solution of the stellar structure equations, and vice versa, it is
necessary to alternately solve both sets of equations in order to obtain a
mutually consistent solution for a given timestep.  Care is taken to ensure that
the diffusion terms transport, but not create or destroy, the total amount of
each species, and moreover, that the creation and destruction of the various
isotopes by the nuclear terms balance appropriately.  Diffusion coefficients are
calculated using the spline fits to the collision integrals for a screened
Coulomb potential given by \citet{ppf86}.

We opted to retain the practice of solving the four stellar structure equations
only in the innermost 99\% of the total mass of the model star, and deriving
the boundary conditions at that point through Runge-Kutta integrations of three
of these equations, assuming that the integrated luminosity is constant in the
surface layers (for a detailed description see \citealt{vee08}; section 2.2).
With this approach, it is possible to follow the large radial variations in the
thermodyanmic quantities in the outer 1\% of the mass, as well as changes in the
size of the convective envelope (especially when it contains $<< 1$\% of the
star's mass), very accurately by using small integration steps.  It is well
known that gravitational settling causes the surface abundances to change the
most rapidly when the convective envelope is the thinnest (see, e.g.,
\citealt{pm91}).  Surface convection zones having fractional masses
$\Delta{\cal M_{\rm SCZ}}/{\cal M_*} =
1 - {{\cal M}_{\rm BCZ}}/{\cal M_*} > 2 \times 10^{-12}$, where
$\Delta\cal M_{\rm SCZ}$ is the mass of the surface convection zone, ${{\cal
M}_{\rm BCZ}}$ is the mass interior to the base of the surface convection zone,
and ${\cal M_*}$ is the total stellar mass, are resolved by the Victoria code.
If the convective-radiative boundary shrinks to the point where the
fractional convective envelope mass falls below this value, the convective
envelope is assumed to consist of a single shell having a fractional mass of
$2 \times 10^{-12}$.

When the outer convection zone contains $< 1$\% of the total stellar mass,
an additional envelope integration (i.e., besides those needed to formulate
the boundary conditions) must be performed at the end of each iteration
of the Henyey solution scheme in order to determine the chemical abundance
profiles in the entire radiative region of the stellar model.  In practice,   
such integrations, for the $\log\,g, \teff$, and surface abundance properties
of the stellar model, are necessary only for the portions of the tracks for
$\sim 0.8\,{{\cal M}_\odot}$, metal-poor stars from somewhat below the turnoff
until approximately the middle of the subgiant branch.  During the subsequent
evolution, the convective envelope deepens rapidly and the radiative-convective
boundary moves into the ``interior" part of the stellar model (where all four
stellar structure equations are solved).  Thus, our procedure is computationally
more efficient than the alternative of solving for the entire structure from
${\cal M}=0$ to ${\cal M}={{\cal M}_*}$ using, e.g., the Henyey technique.
In practice, slightly deeper envelope integrations are performed so that one 
can deal with the case in which the fractional mass in the convective envelope
changes from just over 1\% to just under 1\% in a single timestep.

As already mentioned, uninhibited diffusion in the surface layers of Pop.~II
stars is ruled out by spectroscopic observations of turnoff and lower-RGB stars
in GCs.  Indeed, the hot end of the Li abundance ``Spite plateau"
(\citealt{ss82}) cannot be reproduced by diffusive models {\it unless} extra
mixing below envelope convection zones is taken into account (\citealt{rmr02}).
On the assumption that turbulence is the most important process that competes
with atomic diffusion, \citet[also see \citealt{rmr01}]{rmt00} devised a simple
expression for the turbulent diffusion coefficient in terms of the atomic
diffusion coefficient of helium at a density $\rho_0 = \rho(T_0)$ multiplied
by $c(\rho_0/\rho)^n$, where the constant $c$, the exponent $n$, and the
reference temperature $T_0$ are chosen so as to obtain the best match of the
predicted and observed abundances. 

Although we could have treated turbulent mixing in the same way, we discovered,
as the result of a limited exploration of other ways of defining the turbulent
diffusion coefficient, that the expression $$D_{\rm turb} = c(\rho_{\rm
BCZ}/\rho)^3/(1 - {{\cal M}_{\rm BCZ}}/{\cal M_*})^n\ ,$$ where
$c$ is a constant, $\rho_{\rm BCZ}$ is the density at the base of the
surface convection zone, and $\rho$ is the local density, worked at least as
well as the expression given by Richer et al.~(2000).
(Due to the steep power-law dependence of the density ratio, $D_{\rm turb}
<< 1$ in the nuclear-burning regions of the stellar model, which ensures that
the assumed turbulence will not affect the chemical profiles resulting from
nucleosynthesis and gravitational settling.)
For instance, when we adopt 15.0 for the constant and $n=1.5$, a Standard Solar
Model for the GS98 metals mix is predicted to have a surface Li abundance of
$\log N = 1.10$ (the observed value; see the tables provided by GS98), when
3.31 is the assumed initial value.  (This constraint can be satisfied just 
as well by an SSM for the AGS5 solar abundances if $c$ is set to a value of
33.0.)  Furthermore, using exactly the same equation, with the values of the
free parameters from the solar calibration, in the computation of an
evolutionary track for a $0.76 {{\cal M}_\odot}$ model having $Y=0.25$, and
[Fe/H] $=-2.4$, which has a predicted age of 13.5 Gyr at the turnoff, we obtain
surface Li abundances that vary from the assumed initial value of $\log N =
2.58$ (\citealt{cfo03}) to 2.10, which agrees rather well with the measured
Li abundances in the ``Spite plateau" stars.  However, we note that the
primordial Li abundance from the concordance of WMAP and Big Bang
nucleosynthesis (BBN) has recently been revised upwards from 2.58 to 2.72
(\citealt{cfo08}); consequently, a discrepancy $\sim 0.15$ dex remains.
Encouragingly, evolutionary calculations that were carried out for several
different [Fe/H] values (between $-3.0$ and $-1.8$), and a small range in mass
(consistent with turnoff ages from 10 to 14 Gyr), yielded quite similar Li
abundances values along the upper MS and turnoff.

In view of these results, the simple expression given above for the turbulent
diffusion coefficient was adopted in all of the evolutionary computations that
are presented in this study.  We hasten to add, however, that it has been
subjected only to to the aforementioned tests so far and it could lead to
problems when examined more thoroughly.  This is left for future work to
determine.  For the time being, we can be reasonably confident that this choice
does not adversely affect the $\teff$\ scale of our models, which is perhaps the
main concern with respect to the present paper, because VandenBerg et al.~(2010)
have already shown that the predicted temperatures along isochrones for the
GSCX abundances agree very well with those derived by \citet{crm10} for a large
sample of field stars that have $M_V$ values based on {\it Hipparcos} parallaxes
and which span a wide range in [Fe/H].  Although our treatment of extra mixing
below envelope convection zones is very {\it ad hoc}, our results nevertheless
suggest that stellar interior processes (yet to be understood) will eventually
be found to be responsible for the difference between the observed Li abundances
that define the ``Spite plateau" and the primordial abundance from WMAP-BBN.

Besides the incorporation of diffusion and turbulent mixing, recent improvements
to the rates of several nuclear reactions are now included in the Victoria code.
In particular, the rates given by \citet{jdz98} for $^3$He($^3$He,$\,2p$)$^4$He,
by \citet{cbc07} for $^3$He($^4$He,$\,\gamma$)$^7$Be, and by \citet{jms03} for
$^7$Be($p,\,\gamma$)$^8$Be have been adopted, along with the rate of the
important ``bottle-neck" $^{14}$N($p,\,\gamma$)$^{15}$O reaction by
\citet{mfg08}.  As far as He-burning reactions are concerned, we have opted to
use the rates for the triple-$\alpha$ and the $^{12}$C($\alpha,\,\gamma$)$^{16}$O
reactions reported by \citet{fdb05} and \citet{hfk05}, respectively.  For all
nuclear reactions not explicitly mentioned, rates have been taken from the NACRE
compilation (\citealt{aar99}).  Whereas OPAL opacities are used for temperatures
$\gta 10^4$ K, as already noted in \S~\ref{sec:zmix}, complementary low-T
opacities for all of the assumed chemical abundance mixtures have been generated
using the computer code described by \citet{faa05}.  In addition, we have 
implemented the conductive opacities presented by \citet{cpp07}.  Although some
time was spent to install the sophisticated equation-of-state (FreeEOS)
developed by A.~Irwin\footnote{See http://freeeos.sourceforge.net}, which is
undoubtedly superior to our default EOS (see the Appendix in \citealt{vsr00}),
we decided to continue using the latter for this project because it is much
faster (by about a factor of 3) and, more importantly, the impact of this choice
on predicted evolutionary tracks and lifetimes is barely noticeable.  The main
difference is, in fact, a reduction in the predicted age of a star at the RGB
tip by $\lta 1$\% when the FreeEOS is employed, which is commensurate with a
very slight shift in luminosity along the track.

We conclude this section with two final points.  First, the diffusion physics
described above is taken into account only during the evolution from just prior
to reaching the ZAMS until the end of the subgiant branch.  For the remaining
evolution to the RGB tip, the implementation described by VandenBerg (1992) of
the non-Lagrangian method of solving the stellar structure equations that was
developed by \citet{egg71} is used.  This enables one to compute RGB loci for
low-mass red giants very efficiently (typically requiring $\lta 1\ cpu$-minute
per giant-branch track on a 3 GHz PC) without losing any accuracy insofar as
such predictions as the luminosity of the RGB bump and the helium core mass at
at the onset of the helium flash are concerned (see the next section).   The
neglect of diffusive processes during the relatively short lifetimes of red
giants appears to have only minor consequences, judging from the work of
\citet[also see \citealt{hgt10}, who have introduced atomic diffusion with
an approximate treatment of radiative accelerations into the Eggleton
formalism]{mrr10}.

The second point worth noting is that the pressure at $T = \teff$ is determined
by integrating the hydrostatic equation from very small optical depths to the
photospheric value, in conjunction with a scaled $T$--$\tau$ relation (from
\citealt{vp89}) that reproduces the empirical solar atmosphere of \citet{hm74}.
As discussed by \citet{vee08}, boundary pressures which are derived in this
way agree very well with those predicted by scaled, differentially-corrected
MARCS model atmospheres over wide ranges in temperature, gravity, and
metallicity.  Moreover, the resultant $\teff$\ scale is nearly identical with
that derived by \citet{crm10} for field subdwarfs having $-2.0 \lta$ [Fe/H]
$\lta -0.6$ using the infra-red flux method (see VandenBerg et al.~2010).  This
provides empirical support for our adopted treatment of the surface boundary
conditions, as does the finding that the present models are able to match the
morphologies of globular cluster CMDs quite well, except (possibly) at
the lowest metal abundances (see VandenBerg et al.~2010).

\subsection{Comparisons of Stellar Models Computed Using Different Codes}
\label{subsec:compare}

\begin{figure}
\plotone{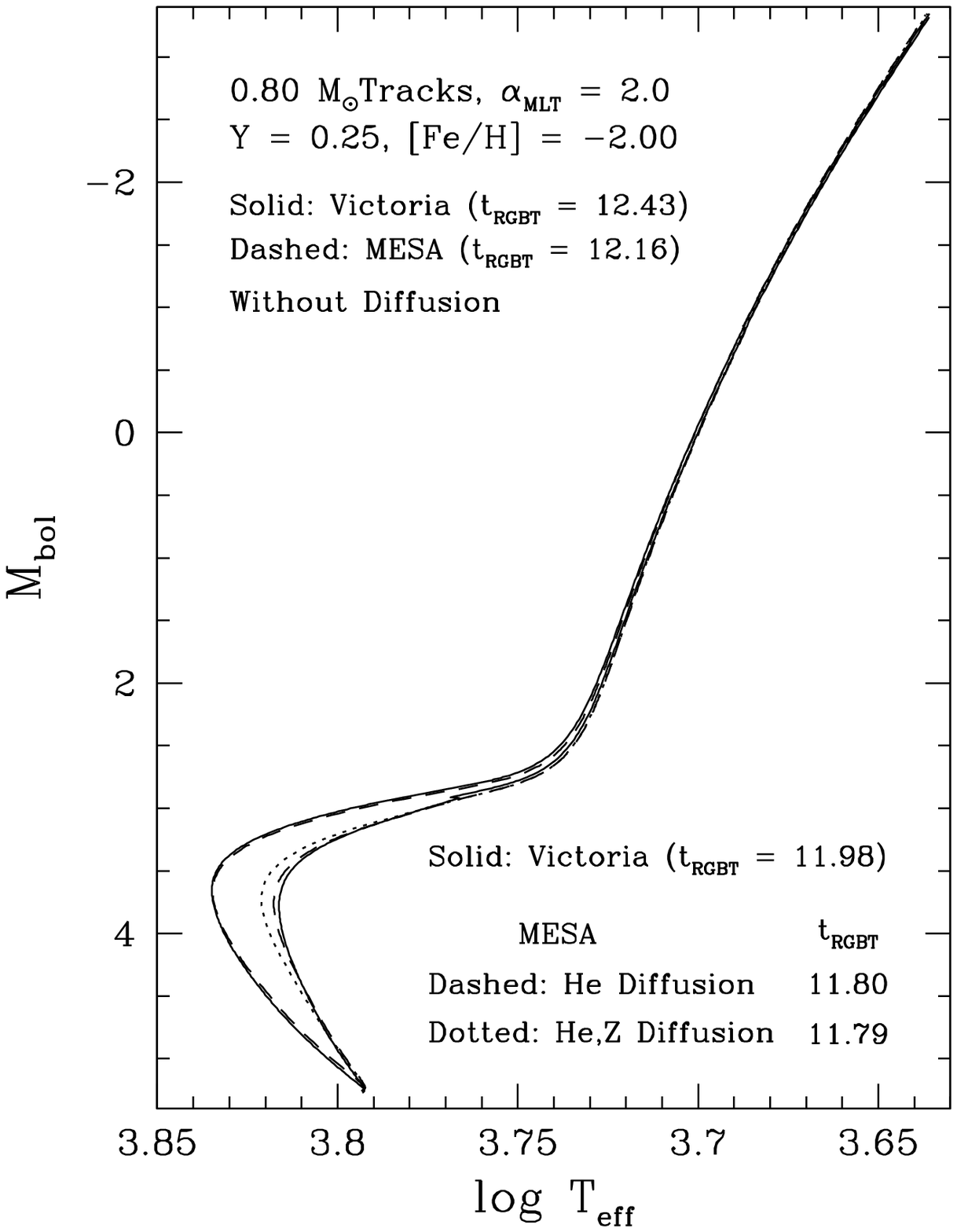}
\caption{Comparison of evolutionary tracks computed using the Victoria and the
MESA codes for the indicated mass and initial chemical abundances if the
gravitational settling of helium is treated (the coolest sequences) and if
diffusive processes are ignored (the hottest tracks).  Predicted ages at the
RGB tip are given in Gyr.  Extra (turbulent) mixing is not considered.}
\label{fig:fig1}
\end{figure}

Figure~\ref{fig:fig1} compares evolutionary tracks that have been computed
using the Victoria and the MESA (version 3168; \citealt{pbd11}) codes for $0.8
{{\cal M}_\odot}$ models having initial values of $Y = 0.25$ and [Fe/H] $= -2.0$
(assuming the GS98 metals mix).  For this comparison, the same value of the
mixing-length parameter, $\alpha_{\rm MLT} = 2.0$, was assumed, and photospheric
pressures were derived by integrating the hydrostatic equation on the assumption
of the \citet[see equation~33]{ks66} $T$--$\tau$ relation.  The tracks with the
coolest turnoffs take diffusive processes into account, while those with the
hottest turnoffs neglect this physics.  Aside from MESA using the \citet{tbl94}
treatment of gravitational settling, both computer programs incorporate the
same, or very similar, basic physics (opacities, nuclear reaction rates,
thermodynamics, convection theory).  Even so, it is remarkable that two
completely independent codes predict nearly identical tracks from the ZAMS to
the RGB tip with, or without, including the settling of helium.  The most
noticeable difference occurs near the turnoffs of the diffusive tracks, which
can probably be attributed to the different treatments of diffusion in the two
codes since there is no significant offset of the turnoff $\teff$s in the tracks
without diffusion.  At the tip of the giant branch, the ages, luminosities, and
helium core masses agree to within $\sim 2$\%, $\delta\log L/L_\odot = 0.015$,
and 0.0007 solar masses, respectively.  Such small variations can easily be
attributed to the uncertainties in the input physics; e.g., they are reduced by
about a factor of two if the equation of state normally used in the Victoria
code is replaced by FreeEOS (see the previous section).

To illustrate the extent to which low-metallicity tracks are affected by metals
diffusion, a track (the dotted curve) has been computed using the same version
of the MESA code, except that the settling of the heavy-elements has also been
treated (i.e., in addition to helium).  As expected, the track evolves to warmer
turnoff temperatures during the core H-burning phase due to the settling of the
metals below the increasingly thin convective envelope.  Because most of those
metals are dredged back into the surface during the evolution along the subgiant
branch when convection penetrates into the interior layers, the track
subsequently merges with the one that allowed only for helium diffusion (the
dashed curve).  However, as already noted, uninhibited diffusion has been ruled
out by spectroscopic observations, which have found little or no difference
between the [Fe/H] values for stars near the turnoff and those on the lower RGB.
If extra mixing is invoked to minimize the variation of the metal abundances
during the evolution between the ZAMS and the turnoff, the resultant track
is shifted to somewhat lower effective temperatures (see the plots provided by
\citealt{rmr02}).

\begin{figure}
\plotone{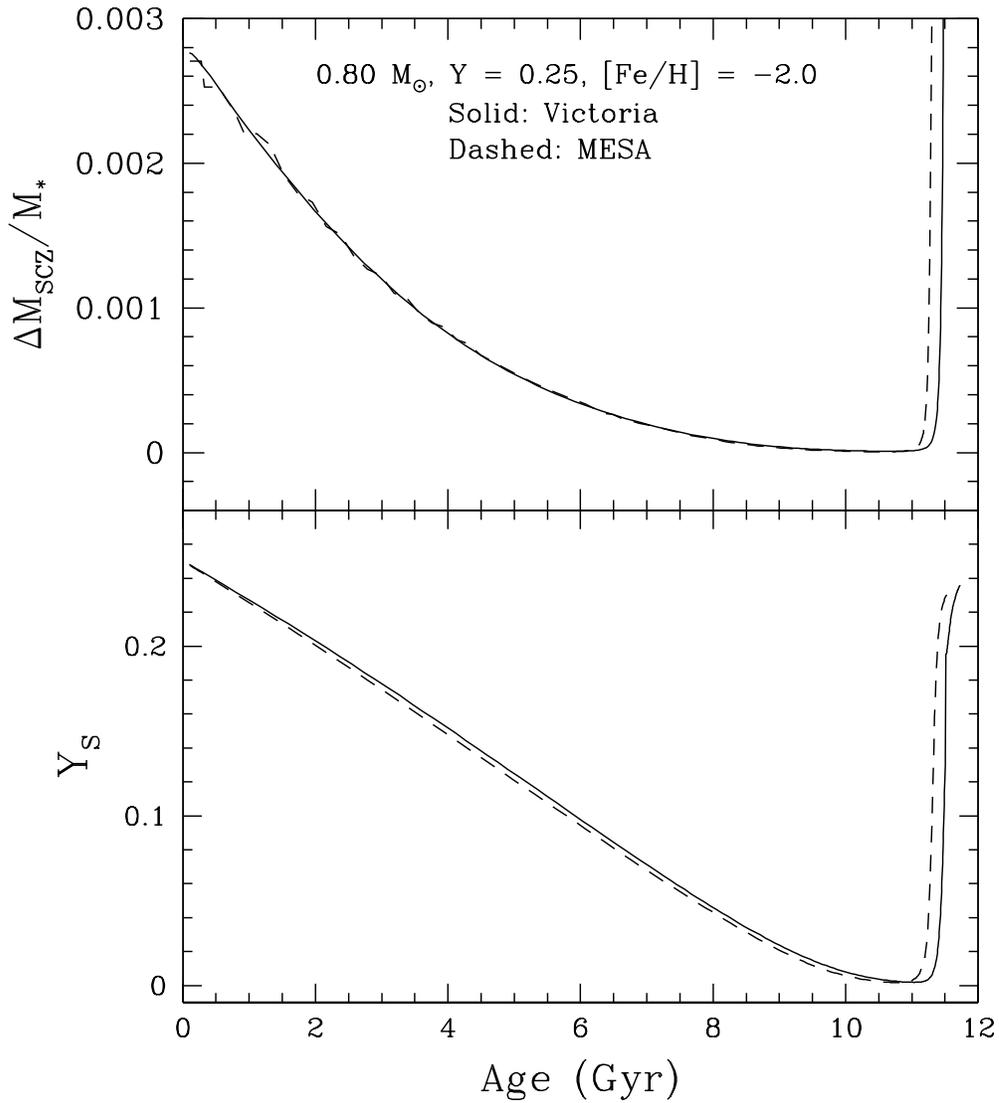}
\caption{The predicted variations with age of the fractional envelope mass in
the surface convection zone (upper panel) and the surface mass-fraction
abundance of helium for the ZAMS to lower RGB portions of the diffusive tracks
shown in the previous figure.  Extra mixing due to turbulence is not taken
into account.}
\label{fig:fig2}
\end{figure}

Given that surface abundance changes are a strong function of the size of the
convective envelope, Figure~\ref{fig:fig2} provides an especially compelling
demonstration that the treatments of diffusion in both codes are very similar.
The upper panel shows how the fraction of the total mass contained within the 
surface convection zone varies with age for the ZAMS to lower-RGB portions of
the diffusive tracks in the previous figure.  The lower panel plots the 
corresponding temporal variation of the surface mass-fraction abundance of 
helium, $Y_{\rm S}$, for the same sequences.  The agreement is clearly very
good.  Note that $Y_{\rm S}$ is predicted to decrease almost to zero when extra
(turbulent) mixing below surface convection zones is ignored, which is the case
in the diffusive tracks considered in Figs.~\ref{fig:fig1} and \ref{fig:fig2}.

\begin{figure}
\plotone{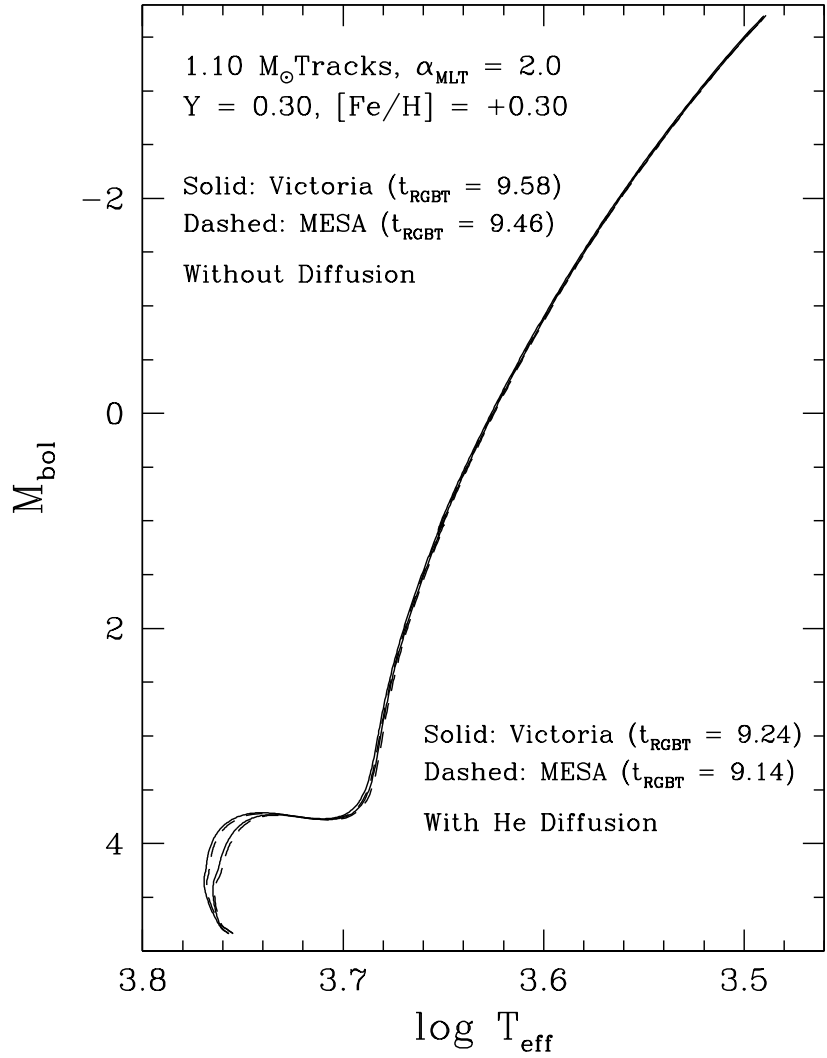}
\caption{As in Fig.~\ref{fig:fig1}, except that tracks have been computed for a
higher mass and higher initial values of $Y$ and [Fe/H], as indicated.
Extra (turbulent) mixing is not treated.}
\label{fig:fig3}
\end{figure}

Although stellar models for super-metal-rich stars are not considered here, we
decided to compare Victoria and MESA tracks for such a case simply to show that
the excellent agreement obtained at low metallicities is also found when
very different masses and chemical abundances are assumed. Figure~\ref{fig:fig3}
is similar to Fig.~\ref{fig:fig1}, except that the evolutionary sequences which
are plotted were generated for a $1.10 {{\cal M}_\odot}$ star, having initial
helium and metal abundances corresponding to $Y = 0.30$ and [Fe/H] $= +0.30$
(assuming the GS98 [$m$/Fe] ratios for all other metals).  At most luminosities,
the predicted $\teff$s agree to within a few Kelvin, with the largest 
differences amounting to $\sim 30$ K near the MS turnoff.  Moreover, at the RGB
tip, the ages agree to within 1.3\%, the luminosities to within $\delta\log
L/L_\odot = 0.011$, and the helium core masses to within 0.0012 solar masses.
This level of consistency is really very satisfactory.

We have also carried out some comparisons with respect to models produced by
the Montreal code (see \citealt{mrr04}; and references therein).  However,
such comparisons are less straightforward to carry out given that, in
particular, the latter use monochromatic opacities to calculate the Rosseland
opacities at each layer in a stellar model and at each evolutionary timestep
(also see \citealt{trm98}).  As noted by Michaud et al., opacity data that are
derived in this way, rather than from interpolations in pre-computed OPAL tables
for a fixed metals mixture (which is the approach used in the Victoria and MESA
codes), have consequences for the solar-calibrated value of $\alpha_{\rm MLT}$
because the variations in the relative abundances of C, N, and O arising
from the operation of the CNO-cycle will modify the opacity profile in stellar
interiors.  Encouragingly, if
%
the value of $\alpha_{\rm MLT}$ is suitably chosen so that ZAMS models
for a given mass and chemical composition have the same convective envelope
mass when generated by either the Victoria or Montreal codes, the predicted
variations of $\Delta{{\cal M}_{\rm SCZ}}/{{\cal M}_*}$ and $Y_{\rm S}$ with
time are nearly the same.  (Plots to illustrate this are not included here
because the level of agreement is comparable to that shown in Fig.~2.) This
indicates that the treatments of diffusive processes which are employed by the
Victoria and Montreal (and MESA) codes lead to very similar effects.  In
fact, isochrones for very metal-poor stars that are obtained from grids of
evolutionary tracks computed using the Montreal code can also be reproduced
rather well by current Victoria computations --- even though the former, but
not the latter, take the diffusion of the metals and radiative accelerations
into account.

\section{Framework for Interpolations in the Model Grids}
\label{sec:interp}

The ultimate goal of the interpolation methods that
we have employed over the years \citep{bv92,bv01,vbd06} is
to produce isochrones, luminosity functions, and isochrone population
functions that represent the underlying canonical models as
accurately as possible.  With this new set of model grids, we add the
ability to interpolate grids of tracks with arbitrary helium abundances
and/or metallicities encompassed by the canonical grids.

The original isochrone interpolation scheme \citep{bv92} relied on
the identification of seven primary equivalent evolutionary phases
(EEPs): 1) the zero-age main sequence (ZAMS), 2) the main-sequence
turn-off (MSTO), 3) the blue hook (BLHK), 4) the Hertzsprung gap
(HZGP), 5) the base of the red-giant branch (BRGB), 6) the giant-branch
pause  (GBPS), and 7) the giant-branch tip (GBTP) on the tracks.
(However, as will be explained below, we have dispensed with the HZGP
EEP; consequently, interpolations in the current models rely on only six
primary EEPs.)  This particular set of EEPs was chosen because they
trace the  mathematical properties of the tracks; some of them also
mark significant evolutionary events.  Except for the ZAMS and GBTP
EEPs, they can be recognized exclusively by the morphology of the
temporal derivative of the effective temperature (i.e., $d(\log  T_{\rm
eff})/d(\log t)$) and all of them can be detected automatically.  Most
importantly, except for some situations in the metal-rich grids when
the CNO cycle begins to emerge as the dominant source of energy
production in the higher mass tracks and blue hook morphology starts
to develop, the age--mass relation for each of the primary EEPs is
monotonic and all of the interpolation relations (e.g., age--mass,
luminosity--mass, or temperature--mass) are very nearly linear.  As
illustrated in Figure 1 of \citet{bv01}, it is easy to recover
monotonic behaviour in the age-mass relations by minor adjustments to
the location of the MSTO and/or BLHK EEPs on the tracks that fall in
this transition region.

When the effects of helium diffusion are included in the model
calculations, a bump feature manifests itself in the subgiant branch
region with an intensity and an extent that is affected by the assumed
chemical abundances.  Problematically, this bump feature migrates across
the HZGP EEP as a function of track mass (though it occurs at a nearly
constant value of $T_{\rm eff}$).  Furthermore, the HZGP EEP as we
originally defined it is an artifact of $pp$-chain processing and it
becomes less significant in  tracks of higher mass. Since it is
difficult to detect the HZGP automatically in tracks that include
helium diffusion and in tracks where the CNO cycle dominates the energy
production, we elected to dispense with it and hence to construct the
interpolation scheme around the remaining six primary EEPs.

The identification of the ZAMS EEP presents an interesting challenge
because the pre-main-sequence (PMS) contraction phases of evolution produce
quite different morphologies depending on the mass of the track and the chemical
composition of the grid.  After considering other possibilities, we decided to 
adopt a ZAMS criterion based on the initial hydrogen content that also takes
into account the iron abundance of the grid as well as the mass of the track
itself:
\begin{displaymath} X_{\rm ZAMS} = X_{\rm init} - 0.0007 - (3.0 + \feh)\times
 (1.667\times 10^{-4} + 4.167\times 10^{-4}({\cal M}/{{\cal M}_\odot} - 0.4)).
 \end{displaymath}
This formulation, which steps past the complicated morphology in the late
contraction phases of the higher mass stellar models, produces monotonic,
smoothly varying age-mass relations connecting the ZAMS EEPs. (For masses
considered in this investigation, the differences between the resultant ZAMS
models and those based on an $L_{\rm grav}/L < 1$\% prescription are quite
minor.  Over the entire range in [Fe/H] and $Y$ for which evolutionary tracks
have been computed for the different choices of the chemical abundances, the
two criteria yield values of $\log L$ and $\log\,T_{\rm eff}$ that differ by
0.0--0.015 and 0.0--0.0010, respectively, nearly independently of the assumed
mass, while age differences range from 0.0--0.017 Gyr, in the case of $1.2
{{\cal M}_\odot}$ models, to $\sim 0.06$ Gyr, in the case of $0.4
{{\cal M}_\odot}$ models.)

It is, in fact, fundamental to the interpolation scheme that age be a
monotonic function of mass for a given primary EEP. In previous papers,
we divided the regions between the primary EEPs into equidistant
secondary EEP intervals via the metric
 \begin{displaymath} \Delta{\cal  D}  =   [1.25(\Delta\log  L)^2  +
  10.0(\Delta\log{T_{\rm eff}})^2]^{1/2}\end{displaymath}
under the assumption that stellar evolution proceeds in a uniform way.
(The right-hand side of the above equation is an arbitrary definition
of the ``distance" along a track.)
It turns out that this assumption is not strictly valid over the range
of chemical abundances and track masses encompassed by the new grids,
particularly on tracks for models in which CNO energy production becomes
competetive with $pp$-chain energy production between the ZAMS and the
MSTO EEP. In the new grids, we employed the distance metric
 \begin{displaymath} \Delta{\cal D}  = [4.0[(\Delta\log  t)^2 +
1.25(\Delta\log L)^2 + 10.0(\Delta\log{T_{\rm eff}})^2]^{1/2}\end{displaymath}
to distribute the secondary EEPs in a way that maintains the monotonicity 
of the age--mass relations.

\subsection{Preparation of the Canonical Grids}
\label{subsec:prep}

To ensure that all astrophysically significant regions of the
HR-diagram would be mapped by the interpolation scheme, we evolved
each track to an age of 30 Gyr (at least) or to the RGB tip, whichever
came first.  A maximum evolutionary age of 30 Gyr ensures that the
isochrones that are interpolated within the canonical grids do not
have any gaps in the point distribution.  However, since evolution
proceeds at a faster rate when, for example, the helium abundance
is increased, and since we chose to use 3-point interpolations between
the grids of tracks, we actually had to increase the maximum
evolutionary age for some tracks in the lower-$Y$ grids to ensure
that the tracks interpolated between the canonical grids would be complete.

Each canonical grid was processed (via the Akima spline, see
\citealt{bv01}) into a set of tracks consisting of secondary EEPs
equidistantly spaced between successive primary EEPs.  The corresponding
regions of each track contain exactly the same number of secondary EEPs.
For tracks that do not evolve as far as the MSTO, we used the central
hydrogen content to match up the secondary EEPs.
The result of all of this is that each canonical grid contains tracks
that have matching complementary secondary EEP distributions across
the range of masses encompassed by them.  For example, the $235^{\rm
th}$ secondary EEP in {\it any} track that has at least that many
points is matched to the same point in every other track in every
other grid.  Grids of tracks at intermediate abundance parameters are
constructed track by track simply by interpolating between matching
secondary EEPs in the canonical grids with abundance parameters that
encompass the interpolants.  For each abundance mixture in this study,
canonical grids were computed for thirteen \feh\ values and for three
different $Y$ values at each iron abundance.   Three-point
interpolation in both \feh\ and $Y$ is used to produce grids at
intermediate abundances.

\subsection{Interpolation tests}
\label{subsec:itest}

\begin{figure}
\plotone{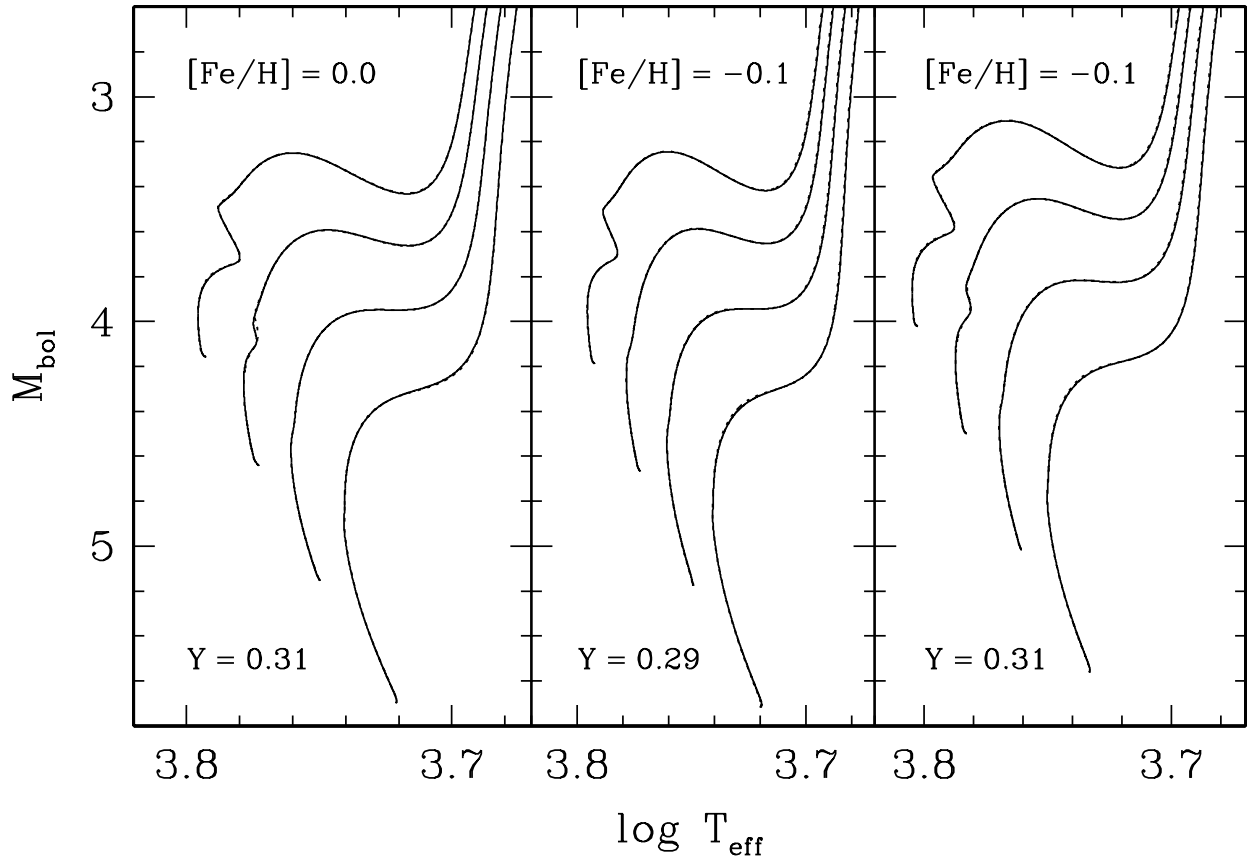}
\caption{1.2, 1.1, 1.0, and 0.9${\cal M}_\sun$ evolutionary tracks from
a special test grid computed directly using the Victoria code are plotted
as {\it solid} lines. Tracks for the same abundance parameters interpolated
from the canonical grids are superimposed as {\it dotted} lines.  The panels,
in the direction from left to right, show examples of interpolations in
$Y$ only, in \feh\ only, and in both  $Y$ and \feh.  In general, the
agreement is astonishingly good, as it is almost impossible to distinguish
between the two sets of tracks.}
\label{fig:fig4}
\end{figure}

\begin{figure}
\plotone{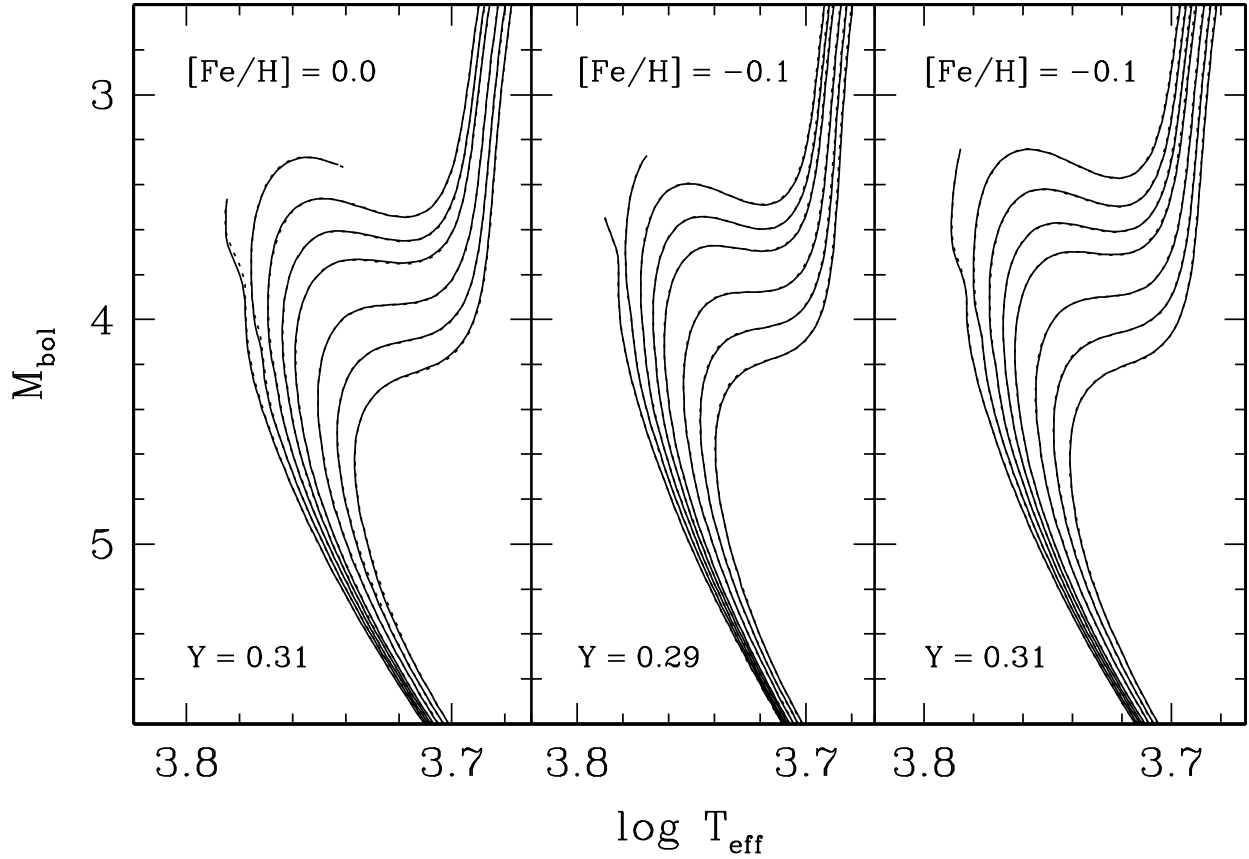}
\caption{4, 5, 6, 7, 8, 10, 12, and 14 Gyr isochrones interpolated directly
from the specially computed test grid are plotted as {\it solid} lines.
Isochrones interpolated from the interpolated tracks are superimposed as
{\it dotted} lines. The experiments shown in each panel correspond to those
illustrated in the previous figure.}
\label{fig:fig5}
\end{figure}

\begin{figure}
\plotone{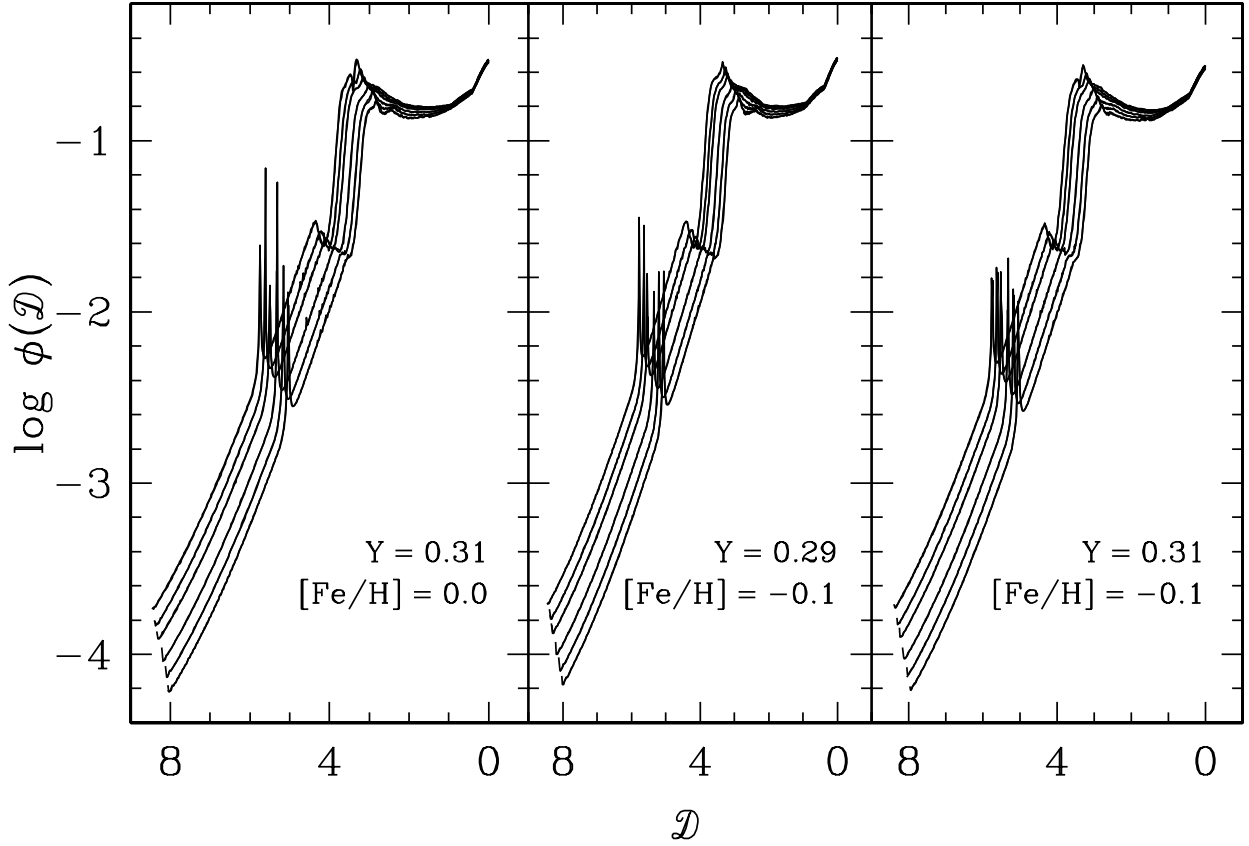}
\caption{Differential isochrone population functions for 4, 5, 6, 7, 8,
10, 12, and 14 Gyr interpolated directly from the special test grid of
tracks are plotted as {\it solid} lines, while those interpolated from the
interpolated grids are superimposed as {\it dotted} lines. The differences
between the two sets of DIPFs are too small to be seen at the scale of
these plots. The experiments shown in each panel correspond to those
illustrated in Fig. 4.  Note that the zero-point of the abscissa coincides
with the position on the H-R diagram of the lowest-mass model that has
been considered, which is for $0.4 {{\cal M}_\odot}$.}
\label{fig:fig6}
\end{figure}

The most obvious test of our interpolation methods is to compare a
grid of tracks that has been interpolated from the canonical grids to
a grid with the same abundance parameters computed directly using the
Victoria code.  The most challenging interpolations occur in grids of
high metallicity and high helium  abundance because these are the ones
in which blue hook morphology manifests itself most strongly in the
intermediate-mass stars.  As illustrated by the examples shown in
Figure~\ref{fig:fig4}, the interpolated tracks match the computed
tracks extremely well --- the only differences which are visible, at
the scale of the plot, appear along the blue hook of the $1.1{\cal M}_\sun$
track in the left-hand panel.  In Figure~\ref{fig:fig5}, we superimpose
isochrones for the ages 4, 5, 6, 7, 8, 10, 12, and 14 Gyr that have
been interpolated from the interpolated tracks over those interpolated
directly from the test set of tracks.  Again, at the scale plotted,
the only noticeable differences are seen along the nascent blue hook
portions of the two youngest isochrones in the left-hand panel.  The most
critical tests of the interpolation scheme are shown in Figure~\ref{fig:fig6},
where we plot the differential isochrone population functions (DIPFs;
see \citealt{bv01}).  Again, there are virtually no detectable
differences at the plotted scale.

\section{The Effects of Chemical Abundance Variations on Stellar Evolutionary
Tracks and Isochrones}
\label{sec:results}

In order to carry out the most thorough examination of the impact of varying
the mix of heavy elements on evolutionary tracks and isochrones, one should use
fully consistent model atmospheres to derive both the surface boundary
conditions for the stellar interior models and the color--$\teff$\ relations
that are needed to transpose the latter to the various CMDs of interest (e.g.,
\citealt{csc04}, \citealt{ssw11}).  Unfortunately, no such atmospheric models
are currently available.  However, varying the abundance of a single metal
should not affect the temperature structure of the atmosphere (i.e., the
$T$--$\tau$ relation) very much (particularly for upper-MS to lower-RGB stars),
and it is well known that colors are much more dependent on temperature than on
metallicity.  Thus, while our results should describe {\it most} of the effects
of chemical abundance variations on the predicted $\teff$\ scale, and thereby
on the model colors, there will be some second-order effects that cannot be
determined until fully consistent ``atmosphere-interior" models are computed
for the assumed abundance mixtures.

The following analyses will mainly consider stellar models for [Fe/H] $= -1.0$
and $-2.0$, given that the majority of the Galactic GCs have metallicities
between these values.  Note that, for most of the metals which are considered,
an increase in the assumed [$m$/Fe] value by 0.4 dex is appreciably larger than
the observed variations within a given GC or from cluster-to-cluster at a given
[Fe/H].  As a result, the predicted effects of such enhancements on H-R diagrams
should be appropriately adjusted to reflect the actual chemical abundance
variations under consideration.  While it seems reasonable to expect that the
net effect of increasing the abundances of two metals at the same time is
equivalent to the sum of the effects of enhancing each metal separately,
especially at low metallicities, it is not known whether this is necessarily
the case.  This and other issues will be addressed in a follow-up study.

\subsection{Increasing the Abundances of Several Metals, in Turn}
\label{subsec:indiv}

\subsubsection{Consequences for Stellar Evolutionary Tracks}
\label{subsubsec:trks}

\begin{figure}
\plotone{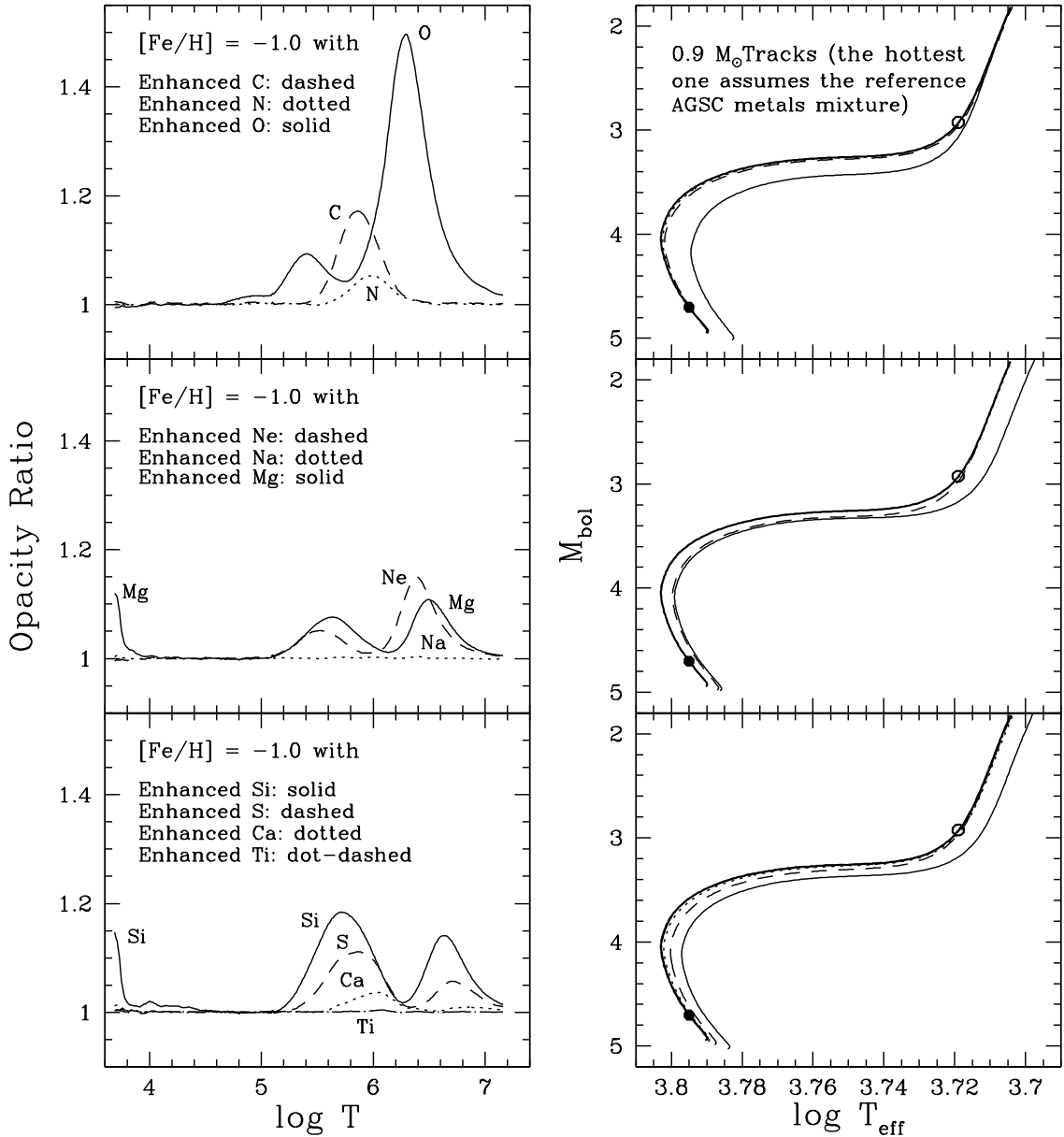}
\caption{The left-hand panels plot ratios of the opacity for each of the
mixtures with an enhanced abundance of a single metal, as indicated, to that
for the reference AGSC mixture, assuming [Fe/H] $= -1.0$.  The temperatures, 
densities, and hydrogen mass-fraction abundances that were assumed in the 
determination of the opacities (by interpolations in tables) correspond to the
surface-to-center variations in the model that has been plotted as a filled
circle in the right-hand panels.  (The structural properties of the lower RGB
model represented by the open circle are used to generate the opacity data
that are plotted in the next figure.)  The definitions of the line types that
are given in the left-hand panels also apply to the adjacent H-R diagrams, which
show the impact of the same abundance enhancements on computed tracks between
the ZAMS and the lower RGB.}
\label{fig:fig7}
\end {figure}

We begin our analysis by comparing the opacities for the different 
heavy-element mixtures at the same densities, temperatures, and hydrogen
mass-fraction abundances, where the latter have been taken from the
computed structures of representative stellar models.  The left-hand panels
of Figure~\ref{fig:fig7} plot, as a function of $\log T$, the ratio of the
opacity for each of the ``enhanced" mixtures (e.g., AGxC, AGxN, etc., see
Table~\ref{tab:tab1}) to that for the reference mixture (AGSC).  These opacities
were calculated for the variations of $\rho$ and $X_{\rm H}$ with $T$ that
describe the structure, from the outermost layers to the center, of a somewhat
evolved $0.9 {{\cal M}_\odot}$ model in the AGSC grid having $Y=0.25$ and
[Fe/H] $= -1.0$.  The ``outermost layers" include that part of a stellar
model which is determined by integrating $d\,P/d\,\tau = g/\kappa$
from very small optical depths to the value of $\tau$
where $T = \teff$ in order to extend the evaluation of the opacity ratio to the
lowest temperatures that are normally considered in a stellar model.  The
upper-left panel shows, for instance, that C, N, and O are important sources of
opacity at temperatures above $\log T \sim 5.5$, while they do not contribute
significantly to the opacity at low temperatures.  The differences in the 
maximum value of the opacity ratio, which vary from $\approx 1.05$ in the case
of nitrogen to $\approx 1.50$ in the case of oxygen, primarily reflect the 
differences in the abundances of these elements.  From the additional panels
along the left-hand side, one sees that (i) Mg and Si are important contributors
to the opacity at both low and high temperatures, (ii) Ne and S are significant
opacity sources only at relatively high temperatures, and (iii) the contribution
of Na and Ti to the opacity is negligible (due mainly to their low abundances).
Interestingly, despite their low abundances, Na and Ca affect the low-$T$
opacities more than most of the other metals, although the differences are
barely discernible at the plotted scale (at least at [Fe/H] $= -1.0$).

Significant changes to the opacity will affect the structure of a stellar model
and, in turn, the location of its track on the H-R diagram.  In particular,
higher interior opacities will tend to shift an evolutionary sequence to lower
luminosities and effective temperatures, whereas the predicted $\teff$, but not
the luminosity, is dependent on the low-$T$ opacities.  The right-hand panels
of Fig.~\ref{fig:fig7} plot the tracks for the reference AGSC mixture along with
the tracks in which each of the 10 metals from C to Ti is given an enhanced
abundance, in turn, by 0.4 dex. As $Y$ and the $\log\,N$ abundances of
all of the other metals are kept the same, the adopted enhancement will
increase $Z$ slightly, and therefore, the value of $X$ that is assumed in the
computed tracks must be decreased in order that $X+Y+Z = 1.0$. (Note that the
line-type identifications are the same as in the left-hand panels.)

Because oxygen is the most abundant of all of the heavy elements and because it,
like C and N, alters stellar models through both opacity and nucleosynthesis
effects, it has the biggest impact, followed by Si, Mg, Ne, S, and C (in the
order of decreasing influence), with the other metals having little or no
effects.  Because the CNO group, as well as Ne and S, are poor electron donors
at low temperatures, the location of the lower RGB is unaffected by variations
in the abundances of these elements.  However, the giant-branch is shifted to
significantly cooler temperatures (by $\sim 100$ K in lower mass stars having
[Fe/H] $\approx -1.0$) if the abundance of Mg or Si is increased by 0.4 dex
over their abundances in the reference mixture (which already has [$\alpha$/Fe]
$= 0.4$). {\it (It is important to remember that the effects that any element
has on evolutionary tracks and isochrones depend on the total amount of that
element which is present.  The above results apply only to the particular case
under consideration; i.e., the consequences of the same increase in the [$m$/Fe]
value of any metal would be smaller if the reference model assumed [$\alpha$/Fe]
$= 0.0$, or if a lower [Fe/H] value were considered.)}

\begin{figure}
\plotone{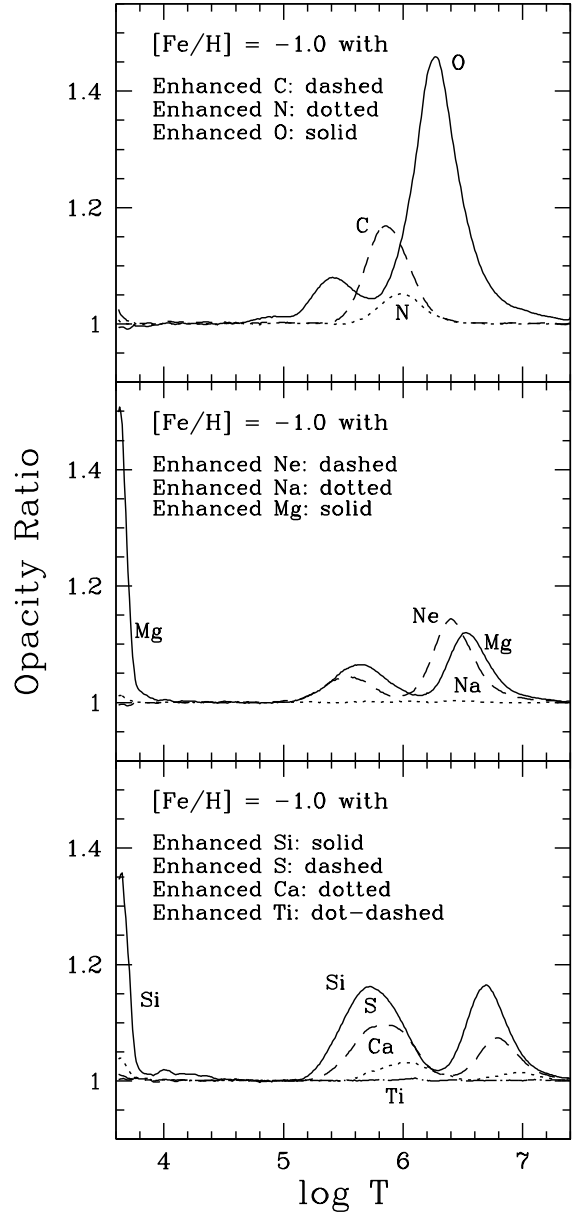}
\caption{As in the left-hand panels of the previous figure, except that the
opacity ratios have been computed for the values of $\rho$, $X_{\rm H}$, and
$T$ from the surface to the center of the lower RGB model that has been plotted
as an open circle in the right-hand panels of that figure.}
\label{fig:fig8}
\end{figure}

The filled and open circles on the reference AGSC track indicate the
locations of the models from which the runs of $\rho$ and $X_{\rm H}$ as a
function of $T$ were taken in order to generate the opacity data that are
plotted in the left-hand panels of Fig.~\ref{fig:fig7} and in
Fig.~\ref{fig:fig8}, respectively.  The two sets of plots look nearly identical,
except at low temperatures, where the opacity ratios reach higher values in the
case of the lower-RGB model because its atmospheric layers are cooler than in 
the MS model.  The differences are especially large when the abundances of Mg
or Si are enhanced, though the increased contributions of C and Ca to the 
low-temperature opacities are also apparent (albeit still very small).  As
shown below in our discussion of upper-RGB stars, their effects on predicted
H-R diagrams are more conspicuous in stars of lower $\teff$.

\begin{figure}
\plotone{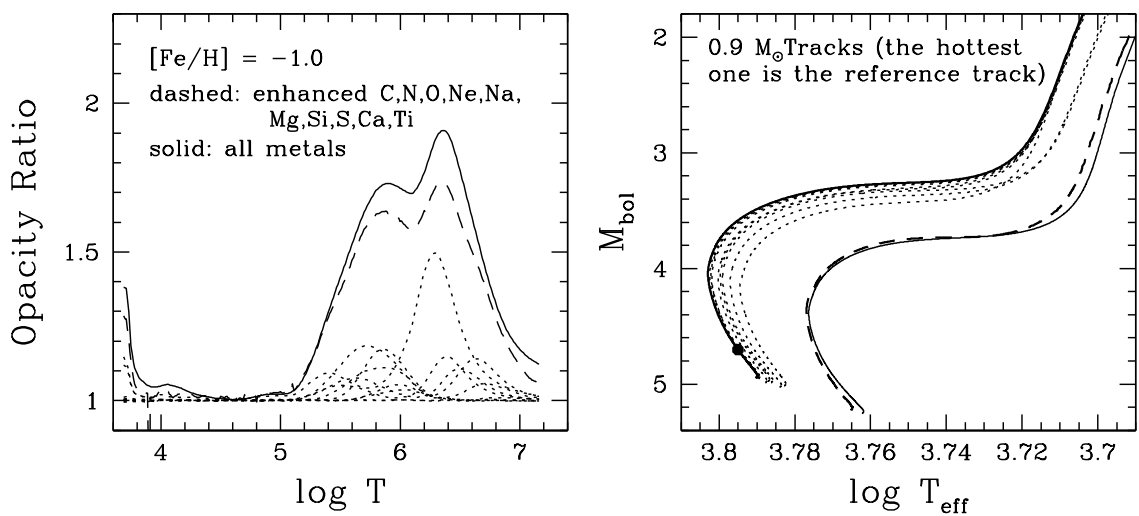}
\caption{As in Fig.~\ref{fig:fig7}, except that all of the loci are plotted
in the same panels as {\it dotted} curves.  The sums of the effects of
increasing the abundance of each metal on the opacity and on evolutionary 
tracks are shown as {\it dashed} curves.  The differences between the latter
and the thin solid curves are (tentatively; see the text) attributed to a 0.4
dex increase in the abundance of iron (primarily), when the abundances of the
10 metals considered here are held constant.  Note that the predicted RGBs for
the tracks with enhanced Mg or enhanced Si overlap one another.  The elements
that have the biggest effects in the vicinity of the turnoff are O and Si,
followed by Mg and Ne.}
\label{fig:fig9}
\end{figure}

Although we do not have opacity data for heavy-element mixtures in which only
the abundance of Fe is increased, it is possible to make an estimate of the
relative importance of this element for opacities and evolutionary sequences if
it is assumed that the effects due
to individual elements are additive.  The dotted
curves in Figure~\ref{fig:fig9} reproduce all of the loci that were originally
plotted in Fig.~\ref{fig:fig7}, while the solid curves are obtained when {\it
all} of the metals are simultaneously given increased abundances by 0.4 dex.
Put in a different way, the thick and thin solid curves in the right-hand panel
represent $0.9 {{\cal M}_\odot}$ tracks for [Fe/H] $= -1.0$ and $-0.6$,
respectively, when both assume the AGSC metals mixture.  Relative to the lower
metallicity case, the models for [Fe/H] $= -0.6$ obviously have higher
abundances of all of the metals by $\delta\log N = 0.4$.  If the differences 
in $\log\teff$ and $M_{\rm bol}$ between the thick solid curve and each of the
dotted curves, as measured along common EEP points, are summed, and then added
to the track for [Fe/H] $= -1.0$, the result is the dashed curve.  It seems
reasonable to attribute the (surprisingly small) offset between this locus and
the thin solid curve to the additional opacity sources that have not yet been
taken into account; namely, Al, P, Cl, Ar, K, Cr, Mn, Fe, and Ni (see
Table~\ref{tab:tab1}).  Of these, iron is the most abundant element, by far;
consequently, the differences between these curves is presumably due mostly
to Fe.

The fraction of the total opacity, and its distribution with temperature, that
is contributed primarily by iron is expected to be approximately the difference
between the solid and the dashed curves in the left-hand panel of
Fig.~\ref{fig:fig9}.  The solid curve plots, as a function of $\log T$, the
ratio of the opacity for [Fe/H] $= -1.0$ to that for [Fe/H] $= -0.6$, assuming
the reference AGSC metals mixture in both cases.  As in previous plots, the
opacities were computed for the values of $T$, $\rho$, and $X_{\rm H}$ that
give the surface-to-center variations of these quantities in the MS model which
has been plotted as a filled circle in the right-hand panel.  The dashed curve,
on the other hand, shows how $1 + \Sigma[\delta$(opacity ratio)] varies with
temperature, where the second term denotes the sum of the dotted curves above
the horizontal line corresponding to the ordinate value of unity.  This plot
suggests that the contribution of iron to the high-temperature opacities is
comparable to those of other abundant metals (like Ne and Si), while it is a
less important source of opacity at low temperatures than Mg or Si.  Indeed,
this is also implied by the right-hand panel: the location of the giant branch
appears to be more dependent on the abundances of Mg and Si than it is of Fe.
However, it is clearly important to check these results by analyzing models in
which the abundance of iron is varied while keeping the abundances of all of the
other metals fixed.  If such an investigation (which will be undertaken in the 
coming months) shows that iron has appreciably larger consequences for opacities
and stellar models than those reported here, one would be forced to conclude
that the repercussions of varying the abundances of individual metals are
not additive.

\begin{figure}
\plotone{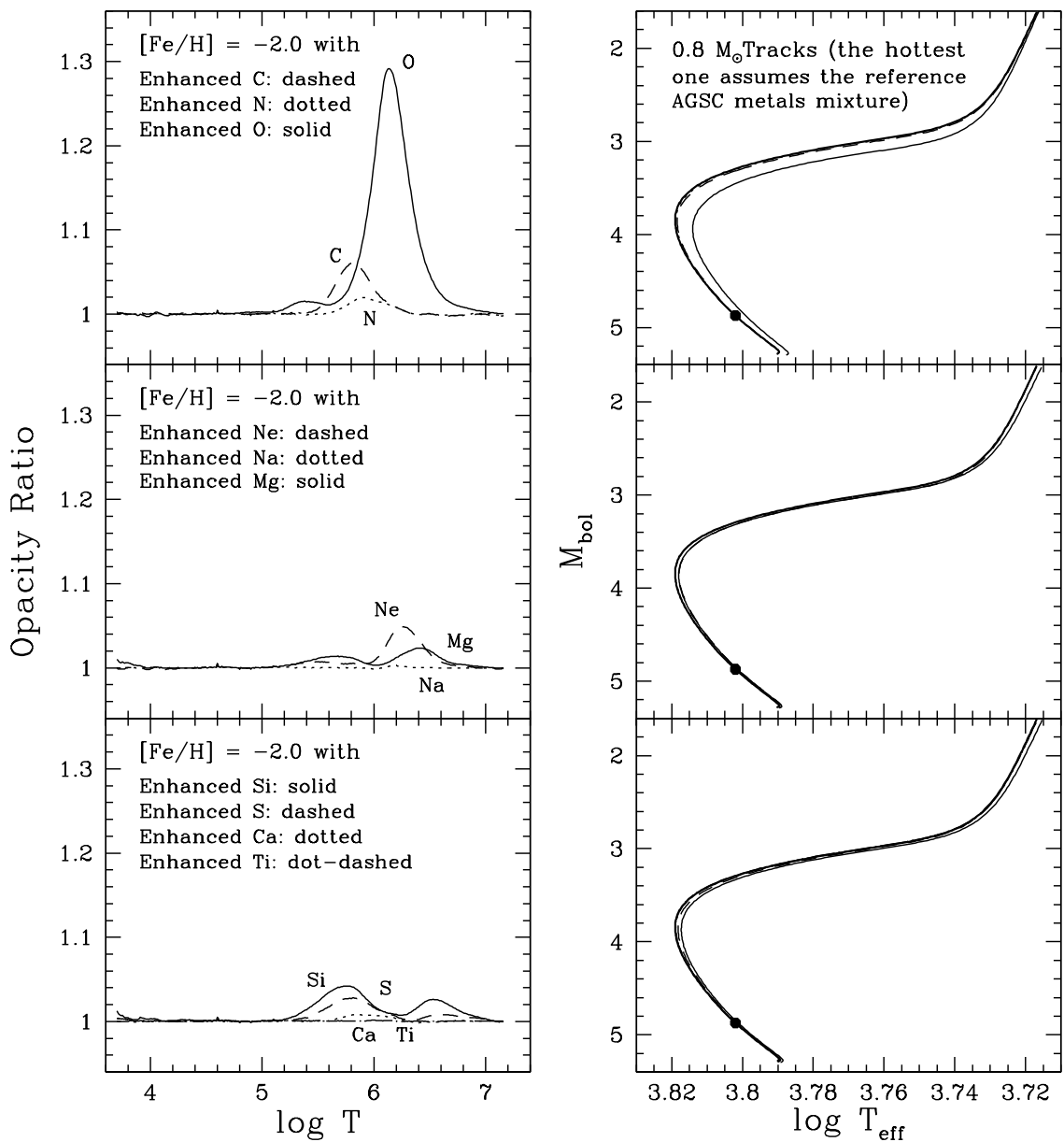}
\caption{As in Fig.~\ref{fig:fig7}, except for [Fe/H] = $-2.0$.  Note that the
range of the ordinate has been reduced, and that the evolutionary tracks for Ne
and Mg overlay one another from the ZAMS until just before the base of the RGB,
at which point they separate.}
\label{fig:fig10}
\end{figure}

Figure~\ref{fig:fig10} plots the same information as in Fig.~\ref{fig:fig7},
except that [Fe/H] $= -2.0$ is assumed.  The individual contributions of the
metals to the opacities are clearly much smaller and the effects of the latter
on computed evolutionary sequences (for $0.8 {{\cal M}_\odot}$ in this case)
are appreciably reduced as well.  Oxygen, neon, magnesium, silcon, sulfur,
and carbon all have some impact on the tracks between the ZAMS and the lower
RGB, while only Mg and Si affect the location of the giant branch.  At lower
metal abundances, the role of the metals (other than CNO) will be of little
significance, unless they have very high [$m$/Fe] ratios.  Because it affects
stellar structures through the operation of the CNO-cycle, oxygen (as well as
C and N, if they are sufficiently abundant) will continue to affect the turnoff
and subgiant morphologies of computed tracks.

\subsubsection{Consequences for Isochrones}
\label{subsubsec:iso}

\begin{figure}
\plotone{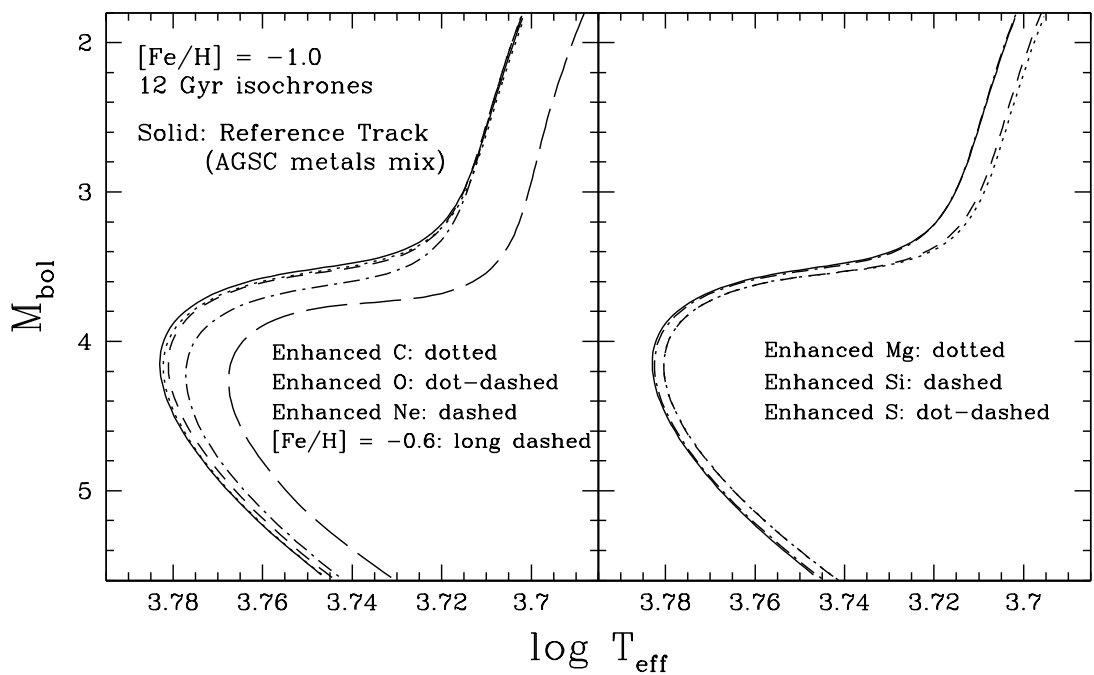}
\caption{Comparison of isochrones for $Y = 0.25$ and [Fe/H] $= -1.0$ with, and
without, 0.4 dex enhancements in the abundances of C, O, Ne, Mg, Si, and S (in
turn).  An isochrone for the same AGSC metals mixture as the {\it solid} curve,
but for [Fe/H] $= -0.6$, has also been plotted to show the impact, relative to
the reference isochrone, of increasing the abundances of all of the metals by
0.4 dex.}
\label{fig:fig11}
\end{figure}

Since most studies of stellar populations make use of isochrones, rather than
evolutionary tracks, we show in Figure~\ref{fig:fig11} how 12 Gyr isochrones
for [Fe/H] $= -1.0$ are altered when the abundances of several metals are
increased by 0.4 dex, in turn.  In fact, it is unneccessary to plot the
isochrones for enhanced abundances of N, Na, Ca, and Ti because they are
essentially identical to that for the reference metals mixture (the solid
curve).  Even those for enhanced C or enhanced S are so close to the reference
isochrone that it is hardly worthwhile to include them in this figure.  Only
the remaining four elements (O, Ne, Mg, and Si) have detectable consequences for
turnoff temperatures (and, in the case of oxygen, turnoff luminosities), and
of these, only Mg and Si affect the location of the RGB.  In order to show
the net effect of simultaneously increasing the abundances of all of the metals
that are listed in Table~\ref{tab:tab1} by 0.4 dex, an isochrone for [Fe/H] $=
-0.60$ (assuming the AGSC metals mix) has also been plotted.  Interestingly, it
appears to have close to the same turnoff luminosity as the oxygen-enhanced
isochrone, which suggests that turnoff-luminosity versus age relations are not
very dependent, if at all, on the abundances of the other metals (at least at
low metallicities).

\begin{figure}
\plotone{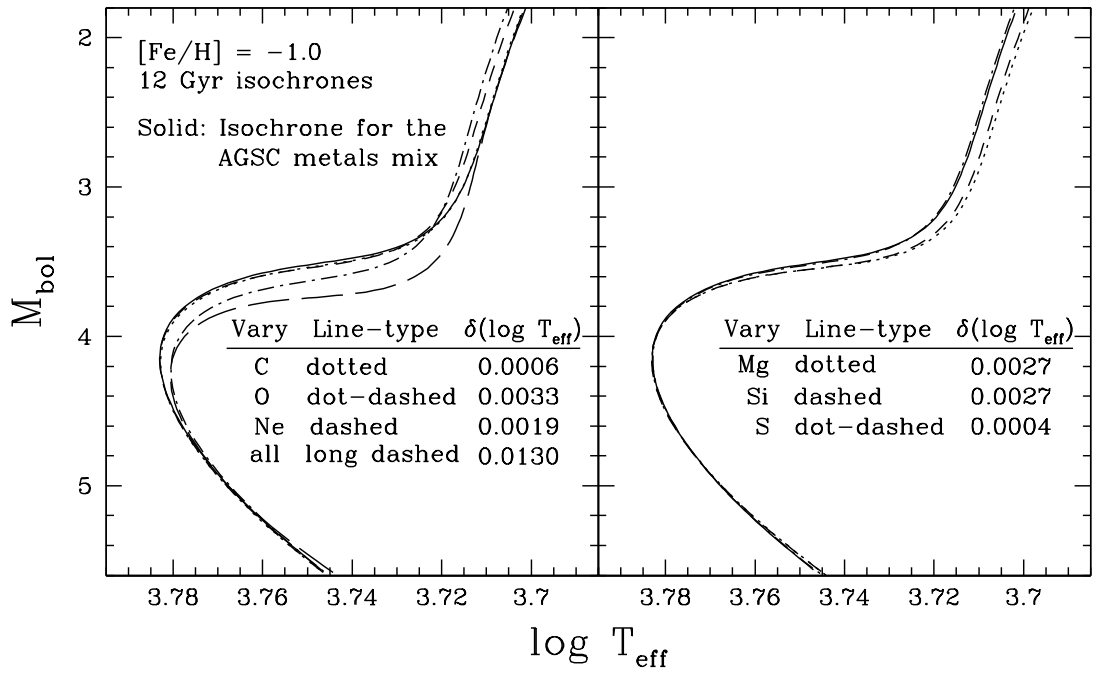}
\caption{As in the previous figure, except that all of the isochrones have 
been shifted horizontally so that they coincide with the reference isochrone
(the {\it solid} curve) at $M_{\rm bol} = 5.0$.  The offsets in $\log\teff$
that have been applied are specified in the legend.}
\label{fig:fig12}
\end{figure}

Indeed, further examination of these isochrones confirms this suspicion.  In
Figure~\ref{fig:fig12}, the same isochrones are replotted after applying 
whatever horizontal adjustments are needed in order to force all of them to have
the same value of $\log\teff$ at $M_{\rm bol}\sim 5$ as the reference isochrone.
(The adopted temperature offsets to achieve this are listed in the legend.)  It
is quite obvious that the isochrones which were computed for the AGSC mixture
with separate 0.4 dex enhancements in the abundances of Ne, Mg, Si, and S have
essentially identical turnoff luminosities as the reference AGSC isochrone.
Only the isochrones for higher abundances of carbon (to a minor extent) and
and oxygen have fainter turnoffs (by $\delta\,M_{\rm bol} \approx 0.02$ and 
0.08 mag, respectively).  As the sum of the latter is almost enough to explain
the even fainter turnoff (by 0.11 mag) of the isochrone for [Fe/H] $= -0.6$,
we infer that the effect of the increased iron abundance on the turnoff
luminosity at a fixed age continues to be insignificant up to metallicities
somewhat greater than [Fe/H] $= -1.0$.

Because O, Ne, Mg, and Si affect the difference in $\teff$ (and hence color)
between the turnoff and the lower RGB {\it at a fixed age}, the use of such a
diagnostic to determine the relative ages of star clusters having very similar
[Fe/H] values (e.g., VandenBerg et al.~1990) will yield reliable results only
if the cluster-to-cluster differences in the abundances of these elements are
small, or if the effects of such differences are taken into account.  It is
still a very worthwhile exercise to compare the locations of the giant branches
of two or more clusters on a CMD, after their turnoffs have been superimposed,
but such comparisons are clearly complicated by the fact that both age and
chemical abundance differences can affect the so-called ``horizontal method of
determining relative ages".  Nevertheless, by considering this and additional
CMD constraints (e.g., the slope of the RGB, the difference in magnitude between
the HB and the turnoff, etc.), it should be possible (at least in principle) to
obtain fully consistent interpretations of the observed photometric data.
 
As discussed above, the reduction in the turnoff luminosity of an isochrone for
a fixed age that occurs when the abundances of all of the metals are enhanced by
0.4 dex appears to be due almost entirely to the increase in the oxygen
abundance.  (Fig.~\ref{fig:fig12} considered the specific case of 12 Gyr
isochrones when the adopted [Fe/H] value is increased from $-1.0$ to $-0.6$ and
the metals have the AGSC [$m$/Fe] ratios.)  This is not too surprising in view
of the fact that, besides being the dominant source of opacity in stellar
interiors (see Figs.~\ref{fig:fig7}--\ref{fig:fig9}), oxygen constitutes most
of the total C$+$N$+$O abundance that governs the rate of the CNO cycle.  It was
shown some time ago (VandenBerg 1992) that, at low $Z$, the impact of higher
oxygen abundances on age-luminosity relations is due much more to the increased 
importance of the CNO cycle than to the concomitant increase in opacity.  The
updated stellar models reported here affirm and reinforce that finding by 
demonstrating that it also applies at higher metallicities than those
considered previously.

\begin{figure}
\plotone{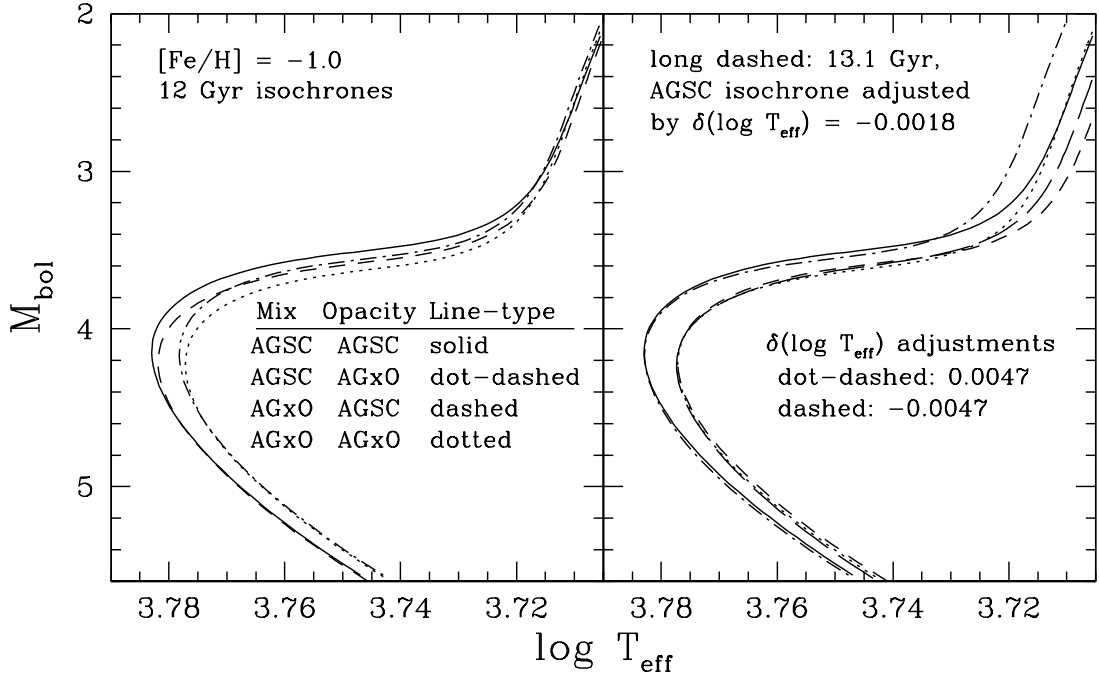}
\caption{The left-hand panel compares isochrones for the AGSC and AGxO cases
({\it solid} and {\it dotted} curves, respectively), with those in which the
opacities that were used to generate the models were not consistent with the
assumed heavy-element mixtures.  To be specific, the {\it dot-dashed} isochrone
assumes the AGSC metals mix and the AGxO opacities, while the {\it dashed}
isochrone assumes the AGxO mixture and the AGSC opacities.  The right-hand panel
shows that, when the {\it dot-dashed} and {\it dashed} isochrones are shifted
in temperature by the amounts indicated, they have very close to the same 
turnoff luminosities as the {\it solid} and {\it dotted} isochrones,
respectively.}
\label{fig:fig13}
\end{figure}

Figure~\ref{fig:fig13} plots 12 Gyr isochrones for [Fe/H] $= -1.0$ in which
the effects of the heavy-element mixtures and the opacities have been decoupled.
The solid and the dotted curves represent, in turn, the isochrones for the AGSC
and the AGxO mixtures wherein the physics has been consistently treated. (Recall
that the AGxO mixture assumes exactly the same number abundances of the metals
as the AGSC mix except for a 0.4 dex increase in [O/Fe].)  On the other hand,
isochrones derived from tracks that assume the AGSC metals mixture but the
AGxO opacities, or the AGxO chemical abundances and the AGSC opacities, are
shown as dot-dashed and dashed curves, respectively.  In the right-hand panel,
the dot-dashed isochrone has been shifted in $\log\teff$ by the amount indicated
in the plot, so as to match the turnoff color of the solid curve.  The two 
isochrones obviously have nearly identical turnoff luminosities, and since they
assume the same AGSC abundances, but different opacities, it follows that the
increased opacity of the AGxO mixture mainly affects the predicted $\teff$
scale.  On the other hand, it is apparent in the left-hand panel that the solid
and the dashed isochrones, which assume different metals mixtures, but the same
opacities, have quite different turnoff luminosities.  Moreover, as shown in the
right-hand panel, the turnoff luminosities of the dashed and dotted curves are
nearly identical.  Thus, it is the high oxygen {\it mixture} and its
consequences for CNO-cycling, rather than its effects on opacities, that is
almost entirely responsible for the difference in the turnoff luminosity between
the solid (AGSC) and the dotted (AGxO) isochrones.  Finally, the long-dashed
isochrone has been plotted in the right-hand panel to show that a 13.1 Gyr
isochrone for the reference AGSC mixture has the same turnoff luminosity as a 
12.0 Gyr isochrone that is otherwise identical, except for the assumption of a
higher oxygen abundance (specifically [O/Fe] $= 0.4$).  This agrees well with
previous determinations of the dependence of turnoff ages on the abundance of
oxygen (e.g., VandenBerg 1992).

\begin{figure}
\plotone{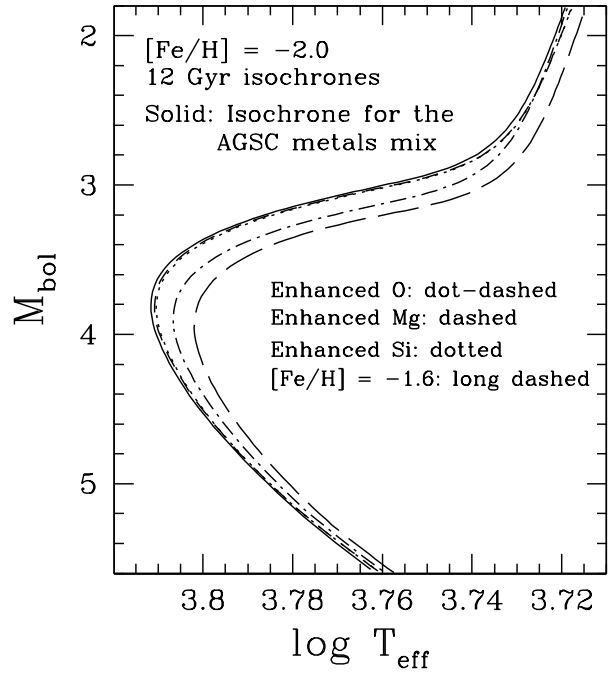}
\caption{Similar to Fig.~\ref{fig:fig12}, except for [Fe/H] $= -2.0$. Isochrones
that allow for 0.4 dex enhancements in the abundances of single metals are
identical with the reference (AGSC) isochrone (the {\it solid} curve), except
in the case of O, Mg, and Si.  An isochrone for [Fe/H] $= -1.6$ has also been
plotted to show the effects of simultaneously increasing the abundances of all
of the metals by 0.4 dex.}
\label{fig:fig14}
\end{figure}

\begin{figure}
\plotone{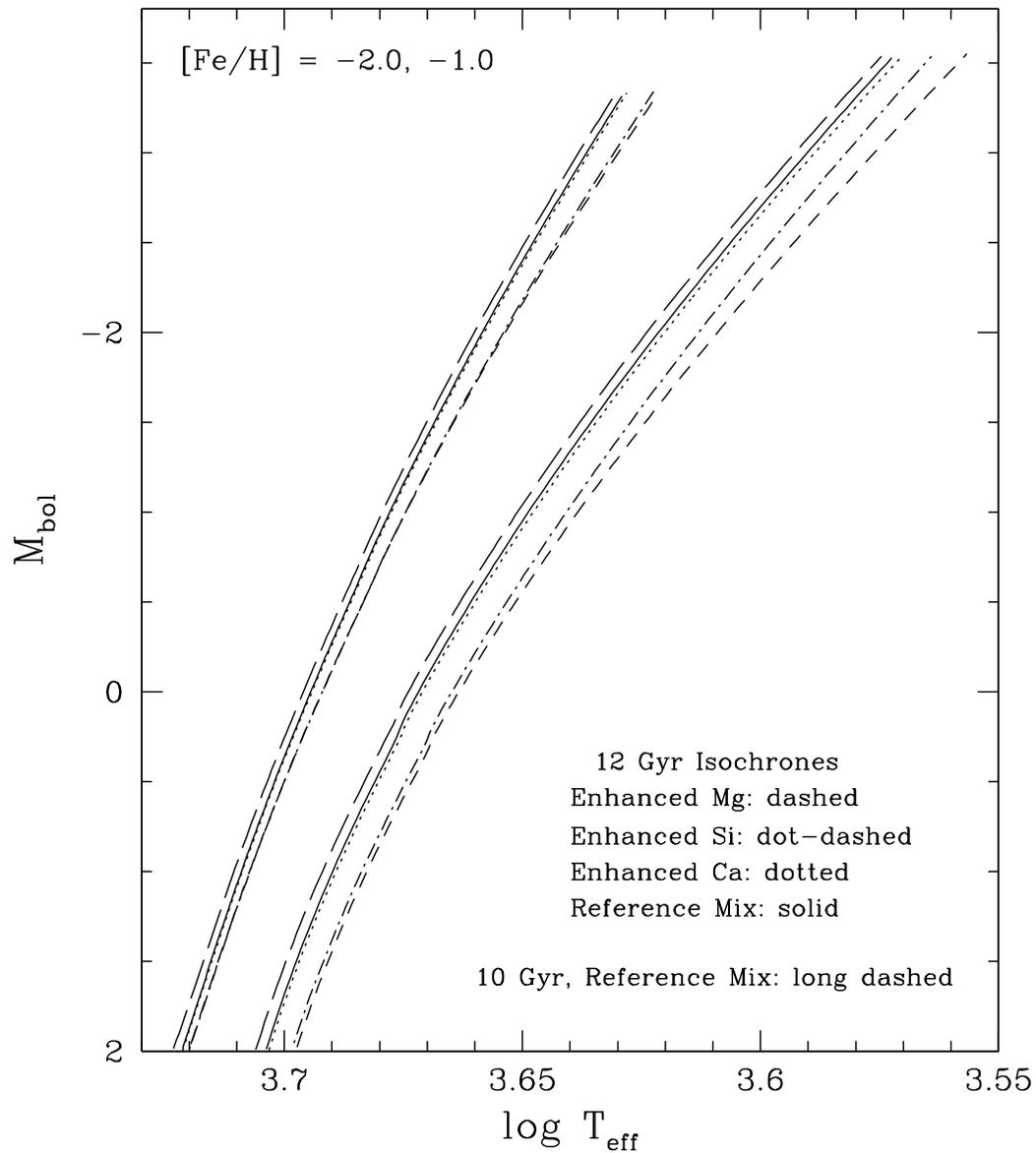}
\caption{Plot of the giant-branch segments of 12 Gyr isochrones for [Fe/H]
$= -2.0$ (the hotter RGBs) and $-1.0$ when enhanced abundances of Mg, Si, and
Ca are assumed, in turn (as indicated).  The {\it solid} and the {\it dashed}
loci represent 12 and 10 Gyr isochrones, respectively, for the reference AGSC
metals mixture.}
\label{fig:fig15}
\end{figure}

\begin{figure}
\plotone{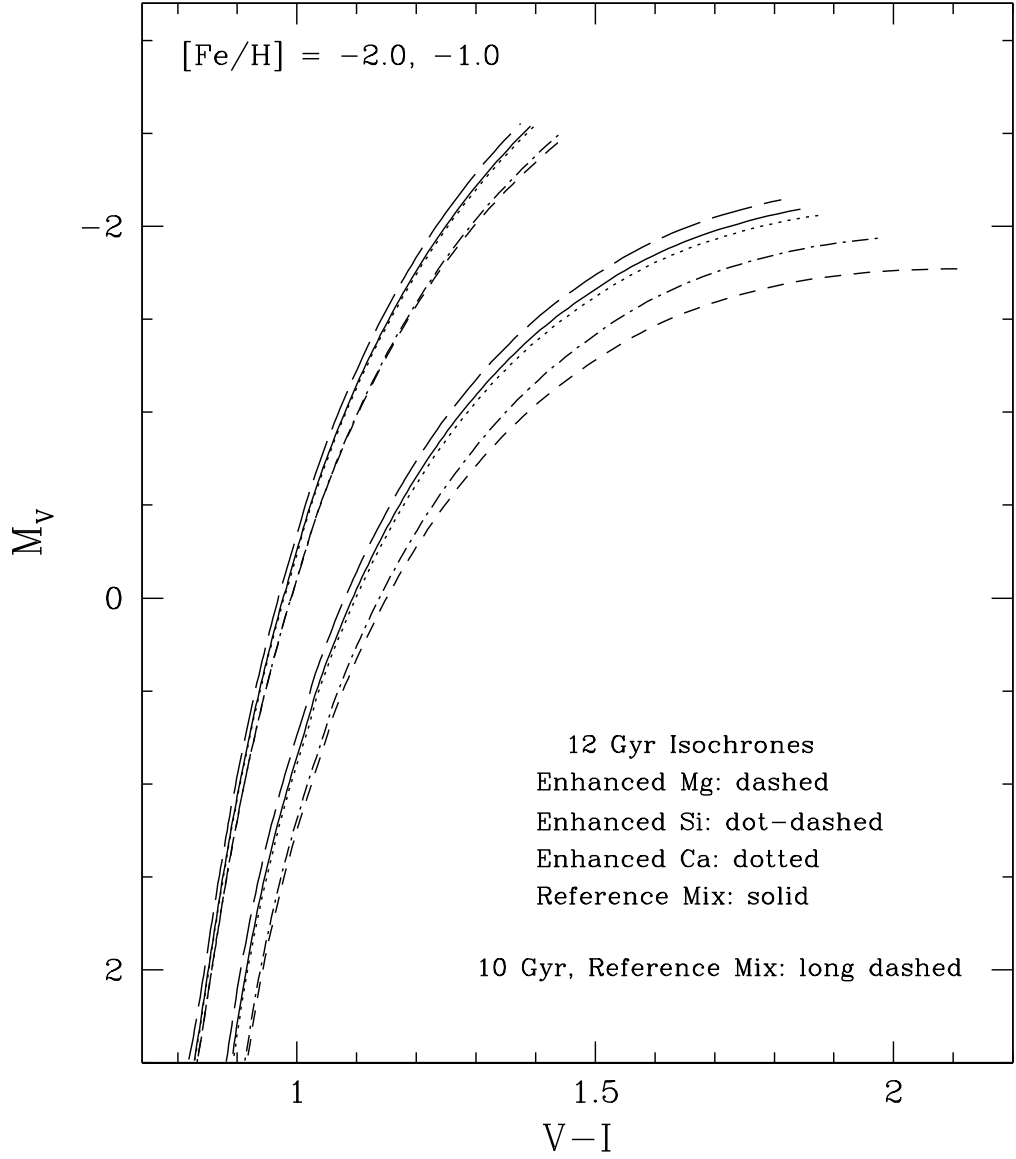}
\caption{As in the previous figure, except that the RGBs of the same isochrones
have been plotted on the $(V-I)\,M_V$ diagram.}
\label{fig:fig16}
\end{figure}

Figure~\ref{fig:fig14} is similar to Fig.~\ref{fig:fig12}, except that it
contains a number of isochrones for [Fe/H] $= -2.0$ instead of $-1.0$.  At the
lower metallicity, 0.4 dex enhancements in the abundances of most of the metals
are inconsequential.  The only elements which do have an impact are O, Mg, and
Si, and of these, only oxygen abundance variations have significant effects on
turnoff-luminosity versus age relations, the difference in $\teff$ between the
turnoff and the lower RGB, etc.  Enhanced Mg or Si abundances mainly cause a
small shift in the respective isochrone to cooler temperatures, but the effects
are too small to be observationally detectable --- at least in the region of the
H-R diagram that has been plotted.  They do become more prominent along the 
upper RGB, as shown in Figure~\ref{fig:fig15}, which indicates that the shift
in $\teff$ resulting from increased abundances of Mg or Si are about a factor
of two higher near the RGB tip than near the base of the giant branch.
On a color-magnitude diagram, the shift in color can be a much stronger 
function of luminosity, depending on the photometric indices that are used.
For instance, Figure~\ref{fig:fig16} shows that, when the same isochrones are
plotted on the $(V-I)\,M_V$ diagram, the isochrones span a range in color near
the tip of the RGB that is $\gta 5$ times larger than at $M_V \sim
2$.\footnote{The color-$\teff$ relations that were used in creating this plot
were derived by L.~Casagrande (see VandenBerg et al.~(2010, section 2) from
synthetic spectra based on the latest MARCS model atmospheres (\citealt{gee08}).
The latter were computed for the AGS5 metals mixture, but with enhanced
abundances of the $\alpha$-elements at low metallicities; specifically,
[$\alpha$/Fe] was assumed to increase linearly from 0.0 at [Fe/H] $= 0.0$ to
$+0.4$ at [Fe/H] $= -1.0$, and to have a constant value of $+0.4$ dex at lower
iron abundances.  Those transformations for the standard $UBV(RI)_CJHK_S$
system, together with similar relations for several {\it HST} ACS and WFC3
filter bandpasses will be the subject of a forthcoming paper by L.~Casagrande
and D.~A.~VandenBerg.}

Interestingly, enhanced Ca has more of an effect on the location of the RGB
than all of the other metals considered here (except Mg and Si), despite its
low abundance.  Indeed, because isochrones for enhanced C, N, O, Ne, Na, S, or
Ti barely deviate, if at all, from the isochrone for the reference mixture,
they have not been plotted in Figs.~\ref{fig:fig14} and~\ref{fig:fig15}.  (Of
course, the very slight shift arising from increased Ca abundances is of no
significance either.)  Note that the effects of varying the abundances of Mg or
Si are considerably larger than those connected with a 2 Gyr change in age (the
difference between the solid and long-dashed curves).  Although it is well
known that the giant branch is much more dependent on metallicity than on age,
it has heretofore not been fully appreciated that Mg and Si are of comparable
importance (at least) as Fe in determining the temperature (and color) of
the RGB, and that the contribution of the other metals is
insignificant.\footnote{Dotter et al.~(2007) certainly
recognized the importance of determining accurate
abundances for Mg and Si, but they found it ``hard to differentiate between
the importance of low- and high-T opacities on stellar models for these two
elements".}  This has important ramifications for, e.g., those
studies which use the differences in the colors of stars that populate the upper
RGBs of very distant systems to derive ages and, in turn, the star formation
histories of those objects.  If there are star-to-star differences in the
abundances of Mg and/or Si, the ages that are inferred  from models which allow
only for different [Fe/H] (and perhaps [$\alpha$/Fe]) values will not be
correct.  In order for such studies, or any investigations that rely on a
detailed understanding of giant stars, to have credible results, accurate and
precise spectroscopic estimates of [Mg/Fe], [Si/Fe], and [Fe/H] will be
required, and the effects of variations in these quantities taken into account.

\subsection{Some Implications of Varying the Solar Metals Mixture}
\label{subsec:solmix}

\begin{figure}
\plotone{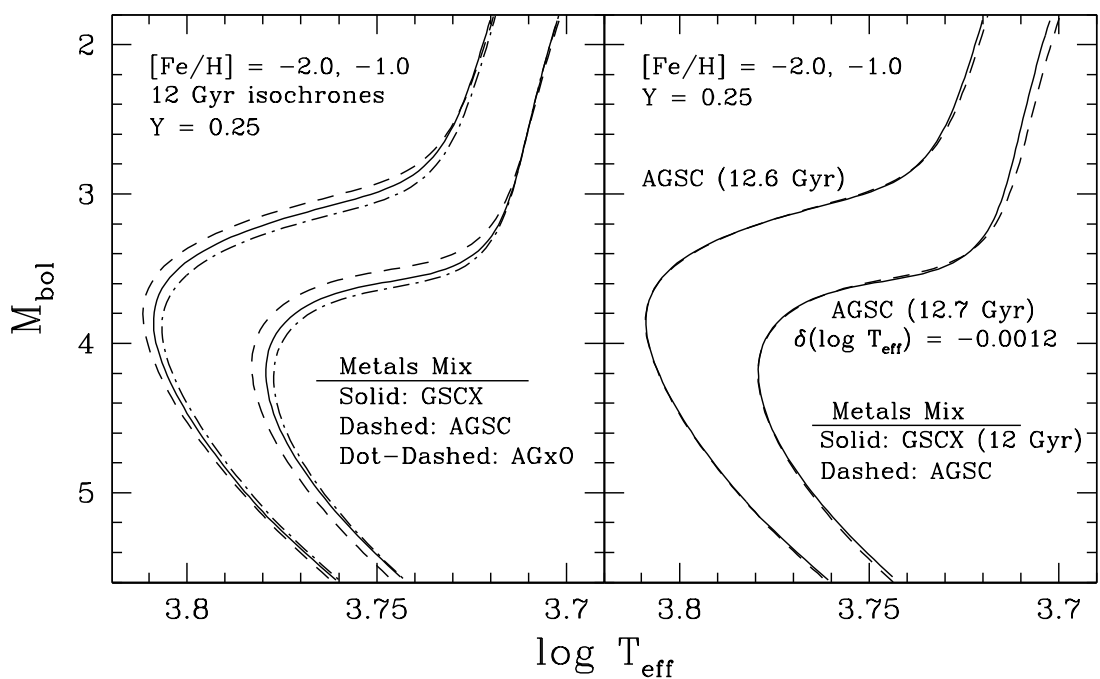}
\caption{The left-hand panel compares AGSC, GSCX, and AGxO isochrones for an
age of 12 Gyr and the indicated $Y$ and [Fe/H] values.  The right-hand panel
shows that, in order to reproduce the turnoff luminosities of 12 Gyr GSCX
isochrones, the AGSC isochrones must have higher ages by $\approx 5$\%.  The
predicted temperatures along the latter have been adjusted by the indicated
amounts to show that they provide an excellent match to the GSCX isochrones
in the vicinity of the turnoff.}
\label{fig:fig17}
\end{figure}

The results discussed above are based on models that assume the AGS5 solar
abundances (see Table~\ref{tab:tab1}) with the additional [$m$/Fe] adjustments
(mainly for the $\alpha$-elements) that were determined for metal-deficient 
stars by C04.  The resultant AGSC models were compared with those obtained when
several of the metals were given further abundance increases, in turn, by 0.4
dex.  It is reasonable to expect that the effects of the latter will, in a
differential sense, be largely independent of the base solar mixture that was
intially assumed.  However, how different are the AGSC isochrones from their
counterparts for the GS98 solar abundances (which we have labelled ``GSCX") in
an absolute sense?  The answer to this question is revealed in the left-hand
panel of Figure~\ref{fig:fig17}, which shows that, at the same values of $Y$,
[Fe/H], and age, the GSCX isochrones have fainter turnoff luminosities.  This
is not too surprising since there is about a 0.2 dex difference in the total
C$+$N$+$O abundance between the AGS5 and GS98 solar mixtures, with the latter
having the higher value of [CNO/H].  It is a well established result that 
higher CNO abundances will have the effect of reducing the turnoff luminosity
at a fixed age.  Confirmation of this explanation is provided by the 12 Gyr
AGxO isochrone that has been included in the plot.  It has a higher oxygen
abundance than either the AGSC or GSCX isochrones, by 0.4 dex and $\approx 0.2$
dex respectively, and it has a fainter turnoff and subgiant branch by about the
expected amounts given the differences between the dashed and solid curves.

As shown in the right-hand panel, the uncertainty in the solar CNO abundances
has about a 5\% effect on age determinations.  AGSC isochrones for $Y = 0.25$
and [Fe/H] $\lta -1.0$ predict ages of 12.6--12.7 Gyr at the same turnoff
luminosities as 12.0 Gyr GSCX isochrones for the same chemical abundances. (Note
that the indicated offsets in $\log\teff$ have been applied to the AGSC loci
simply to demonstrate that they do, indeed, reproduce the turnoffs and subgiant
branches of the GSCX isochrones when the aforementioned age differences are
assumed.)  On the other hand, spectroscopic determinations of [O/H] in
metal-poor stars would imply higher values of [O/Fe] in those stars by about
0.2 dex if the solar value of [O/H] that was used to calculate [O/Fe] is taken
from the AGS5 determination instead of GS98.  Hence, it is important to ensure
that the isochrones which are used to derive ages have been computed for the
correct value of [O/Fe], which depends on how this quantity was derived.  In
fact, the models that are compared with observations should be generated for
the observed absolute abundance (i.e., for the measured $\log\,N_i$ value, on
the usual system where $\log\,N_{\rm H} = 12.0$, or the [$m$/H] value) of each
of the most important elements, rather than for such secondary abundance
indicators as [$m$/Fe].

\subsection{Varying the Helium Abundance}
\label{subsec:helium}

\begin{figure}
\plotone{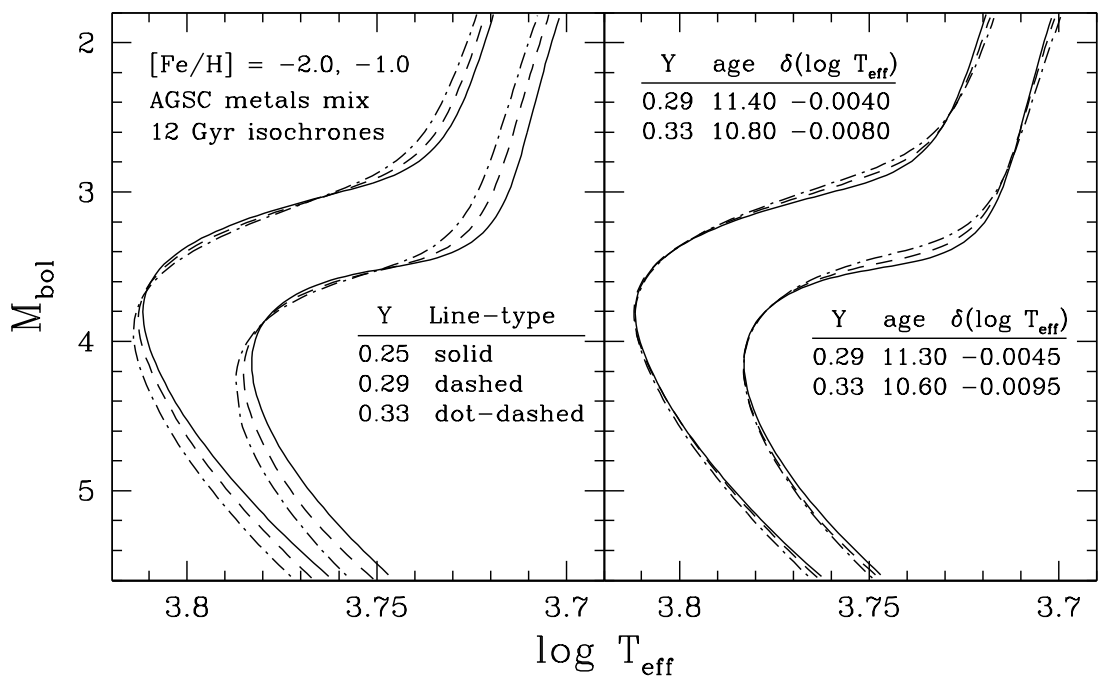}
\caption{The left-hand panel illustrates the effect of varying $Y$ on 12 Gyr
isochrones for [Fe/H] $= -2.0$ and $-1.0$ (assuming the AGSC metals mix).  The
right-hand panel compares 12 Gyr isochrones for $Y = 0.25$ with those for
$Y = 0.29$ and 0.33, assuming those ages for the latter such that all three of
the isochrones for a given [Fe/H] value have the same turnoff luminosities.
The higher $Y$ isochrones were adjusted in $\log\teff$ by the indicated amounts
simply to show that they provide a close match to the turnoff of the {\it solid}
curve.}
\label{fig:fig18}
\end{figure}

Since the model grids reported in this investigation have been computed for
three values of $Y$ (0.25, 0.29, and 0.33) for each adopted [Fe/H] value, it
is worthwhile to briefly discuss the dependence of isochrones on the assumed
helium abundance.  As pointed out by \citet{car81} more than 30 years ago, and
illustrated in the left-hand panel of Figure~\ref{fig:fig18}, the location on
the H-R diagram of the subgiant branch of computed isochrones for a fixed age
and metallicity is nearly independent of $Y$.  Higher $Y$ shifts both the main
sequence and the giant branch to higher effective temperatures and it causes a
significant reduction in the turnoff luminosity at a given age --- or,
equivalently, younger ages at a fixed turnoff luminosity.  This effect is
quantified in the right-hand panel, which shows that a 12.0 Gyr isochrone for
$Y = 0.25$ has the same turnoff $M_{\rm bol}$ as isochrones for 11.4 Gyr, if
$Y = 0.29$, or for 10.8 Gyr, if $Y = 0.33$ (assuming [Fe/H] $= -2.0$ and the
AGSC metals mixture).  From the information that is provided in the figure, one
sees that very similar results are obtained for the [Fe/H] $= -1.0$ isochrones
which have been plotted.

\section{Summary}
\label{sec:sum}

This investigation was carried out primarily to determine the impact on computed
evolutionary tracks (for masses from 0.4 to $1.2 {{\cal M}_\odot}$) and 
isochrones (for ages in the range of 5--14 Gyr) of varying the abundances
of each of several metals (specifically C, N, O, Ne, Na, Mg, Si, S, Ca, and Ti),
in turn, at constant [Fe/H] values.  Indeed, extensive model grids for each of
these cases have been computed for
[Fe/H] values ranging from $-3.0$ to $-0.6$, in 0.2 dex
increments, with $Y = 0.25, 0.29$, and 0.33 at each [Fe/H] value.  All of the
tracks were generated using a significantly improved version of the Victoria
stellar structure code, which now includes a treatment of the gravitational
settling of helium and extra (turbulent) mixing very similar to the methods
employed by \citet{pm91}, as well as the latest nuclear reaction rates and
conductive opacities.  Fully consistent OPAL and low-temperature Rosseland mean
opacities were obtained, and used, for each of the adopted heavy-element
mixtures.  When very similar physics is assumed, the Victoria models are found
to be in excellent agreement with those based on the MESA or the University of
Montreal codes.  Important improvements have also been made to the Regina
interpolation code, which now produces isochrones, luminosity functions, etc.,
from a set of evolutionary tracks very efficiently using an interactive
iterative interface.  Moreover, it may be used to interpolate for arbitrary $Y$
and [Fe/H] values within the ranges for which grids of evolutionary tracks have
been computed.  We have verified that the interpolated grids for intermediate
values of $Y$ and [Fe/H] reproduce those which are computed specifically for
those abundance choices remarkably well, with respect to not only the placement
of the tracks on the H-R diagram, but also the temporal derivatives of the
luminosity and effective temperature along the tracks, which are important for
the calculation of isochrone probability  functions.

Because they are poor electron donors at low temperatures (i.e., they do not
contribute significantly to low-temperature opacities), C, N, O, Ne, and S
do not affect the predicted location of the giant branch.  However, as they are
important sources of opacity at stellar interior conditions, tracks
computed for enhanced abundances of these elements are somewhat cooler and
fainter.  C, N, and O
(especially oxygen, because of its great abundance) have bigger effects on the
turnoff-luminosity versus age relations from isochrones than any other metal
mainly because of their role in the CNO cycle.  Higher CNO abundances imply an
increased importance of the CNO cycle and thereby more centrally concentrated
burning which, in turn, has the ramification that the radius of a star grows
more rapidly and the turnoff is reached sooner.  By contrast, the effects on
isochrones due to changes in the opacity are much less important.

The remaining elements that were considered --- Na, Mg, Si, Ca, and Ti --- are
important opacity sources at both low and high temperatures, but because of
their low abundances, the effects of Na, Ca, and Ti are of no consequence (at
least at low metallicities).  As a result, only Mg and Si (and Fe, which is not
given the same detailed analysis as the other elements) impacts the models for
the RGB phase.  Indeed, our study suggests that the location of the giant
branch on the H-R diagram is a strong function of the abundances of Mg and Si,
seemingly stronger than its dependence the iron content, which would not be too
surprising given that the former are more abundant than the latter, especially
in an $\alpha$-element enhanced metals mixture.  Mg and Si also have significant
consequences for the $\teff$ scale of MS and turnoff stars: only oxygen is more
important for the turnoff phase of evolution.  As opacity effects are largest
at the coolest temperatures, and since O, Ne, and S affect the temperatures of
MS and turnoff  stars but not those of giants, variations in the abundances of
O, Ne, Mg, Si, and S can all affect the difference in temperature (or color)
between the turnoff and the lower RGB, when measured at some luminosity above
the turnoff.  As a result, it is risky to use this separation as a constraint
on, in particular, the relative ages of star clusters having similar iron
abundances (the method first proposed by VandenBerg et al.~1990), unless the
abundances of these 5 metals have been determined and the effects of
cluster-to-cluster differences in them taken into account.   Star-to-star
variations in the abundances of Mg and Si may also be responsible for the
broadening of the RGB in some GCs, and spreads in the abundances of these
metals, plus O, Ne, and S, could potentially affect the width of the MS.

As long as it is clear which solar abundances (e.g., GS98, AGS5, A09) have been
used in deriving the [$m$/Fe] values from the observed $m$/H number-abundance
ratios in metal-poor stars, and the correct [$m$/Fe] values are assumed in the
computation of the isochrones which are fitted to globular cluster CMDs, the 
ages so obtained should be independent of this choice.  Importantly, the turnoff
luminosity versus age relations from isochrones are primarily a function of the
oxygen and helium abundances.  Finally, we note that it is our intention to 
carry out a supplementary study to address a number of unanswered questions.
Besides carrying out some comparisons of the observed widths of the principal
photometric sequences of globular clusters with models that allow for varying
abundances of several metals (notably O, Mg, and Si), the forthcoming work will
isolate and examine the effects of varying the Fe abundance (keeping all other
abundances fixed), and it will determine, for cases of particular interest
(including Ne--Na and Mg--Al anticorrelations), the impact of varying the
abundances of two elements at the same time.  Such work will shed some light
on whether the net effect of varying the abundances of several elements is
equivalent to adding the effects due to each individual element, and it will
tell us, for instance, whether the tracks and isochrones that are computed for
normal Mg and Al abundances reproduce those for reduced Mg and the
significantly enhanced Al abundance that would be expected (because the latter
is initially much less abundant than the former) if the sum of the Mg and Al
abundances is constant.  (It is our intention to make the model grids that will
be computed in the forthcoming study, along with a subset of those generated
for the present investigation, generally available through the Canadian
Astronomical Data Center.  Those with a more immediate need for some of the
computations reported herein are asked to contact the first author.)

\acknowledgements
We are grateful to the referee of this paper, Achim Weiss, for a number of
suggestions that have led to an improved paper.  We also thank Karsten
Brogaard for a number of helpful suggestions on the manuscript.  AD
acknowledges the support of a CITA National Fellowship during the period
of time when he completed most of his contribution to this project.  This
work has also been supported by the Natural Sciences and Engineering Research
Council of Canada through Discovery Grants to DAV and GM.

\clearpage
 
\clearpage

\clearpage

%

\clearpage

\clearpage

\clearpage

\clearpage

\clearpage

\clearpage

\clearpage

\clearpage

\clearpage

\clearpage

\clearpage

\clearpage

\clearpage

\clearpage

\clearpage

\clearpage
\begin{deluxetable}{cccccc}
\tabletypesize{\small}
\tablewidth{250pt}
\tablecaption{Adopted Heavy-Element Mixtures \label{tab:tab1}}
\tablewidth{0pt}
\tablehead{ & \multispan{3}{\hfil $\log N$\tablenotemark{a} \hfil} &
 \multispan{2}{\hfil $\Delta\log N$} \hfil \\
 & \multispan{3}{\hrulefill} & \multispan{2}{\hrulefill} \\
\colhead{Element} & \colhead{A09\tablenotemark{b}} &
 \colhead{GS98\tablenotemark{c}} & \colhead{AGS5\tablenotemark{d}} &
 \colhead{\phantom{$-$}C04\tablenotemark{e}} &
 \colhead{extra\tablenotemark{f}}}
\startdata
C  & 8.43 & 8.52 & 8.39 & \phantom{$-$}0.00 & 0.40 (AGxC) \\
N  & 7.83 & 7.92 & 7.78 & \phantom{$-$}0.00 & 0.40 (AGxN) \\
O  & 8.69 & 8.83 & 8.66 & \phantom{$-$}0.50 & 0.40 (AGxO) \\
Ne & 7.93 & 8.08 & 7.84 & \phantom{$-$}0.30 & 0.40 (AGNe) \\
Na & 6.24 & 6.33 & 6.17 & \phantom{$-$}0.00 & 0.40 (AGNa) \\
Mg & 7.60 & 7.58 & 7.53 & \phantom{$-$}0.30 & 0.40 (AGMg) \\
Al & 6.45 & 6.47 & 6.37 & \phantom{$-$}0.00 & \\
Si & 7.51 & 7.55 & 7.51 & \phantom{$-$}0.40 & 0.40 (AGSi) \\
P  & 5.41 & 5.45 & 5.36 & \phantom{$-$}0.00 & \\
S  & 7.12 & 7.33 & 7.14 & \phantom{$-$}0.30 & 0.40 (AGxS) \\
Cl & 5.50 & 5.50 & 5.50 & \phantom{$-$}0.00 & \\
Ar & 6.40 & 6.40 & 6.18 & \phantom{$-$}0.25 & \\
K  & 5.03 & 5.12 & 5.08 & \phantom{$-$}0.00 & \\
Ca & 6.34 & 6.36 & 6.31 & \phantom{$-$}0.30 & 0.40 (AGCa) \\
Ti & 4.95 & 5.02 & 4.90 & \phantom{$-$}0.25 & 0.40 (AGTi) \\
Cr & 5.64 & 5.67 & 5.64 & $-$0.30 & \\
Mn & 5.43 & 5.39 & 5.39 & $-$0.40 & \\
Fe & 7.50 & 7.50 & 7.45 & \phantom{$-$}0.00 & \\
Ni & 6.22 & 6.25 & 6.23 & \phantom{$-$}0.00 & \\
\enddata
\tablenotetext{a}{On the scale in which $\log N$(H) = 12.0}
\tablenotetext{b}{A09 $=$ Asplund et al.~(2009)}
\tablenotetext{c}{GS98 $=$ Grevesse \& Sauval (1998)}
\tablenotetext{d}{AGS5 $=$ Asplund, Grevesse, \& Sauval (2005)}
\tablenotetext{e}{C04 $=$ Cayrel et al.~(2004)}
\tablenotetext{f}{Separate model grids, with the name identification given
in the parentheses, were generated for the ``AGS5 $+$ C04 $+$ extra" metal
abundances, scaled to many [Fe/H] values}
\end{deluxetable}

\end{document}